\documentclass[12pt]{article}
\usepackage[latin1]{inputenc}

\usepackage{amsmath}
\usepackage{color}
\usepackage{amsfonts}
\usepackage{amssymb}
\usepackage[mathscr]{euscript}
\usepackage{graphicx}
\usepackage{geometry}
\usepackage{amssymb,epsfig,subfigure}
\usepackage{hyperref}
\usepackage{comment}
\usepackage{tabularx}
\usepackage{bm}
\usepackage{euscript}
\usepackage{graphicx}
\usepackage[dvipsnames]{xcolor}
\usepackage{amsfonts}
\usepackage{exscale}
\usepackage{amsbsy}
\usepackage{subfigure}
\usepackage{textcomp}
\usepackage{comment}
\usepackage{hyperref}
\usepackage{slashed}
\usepackage{authblk}
\usepackage{tabularx}
\usepackage{euscript}
\usepackage{graphicx}
\usepackage{color}
\usepackage{amsfonts}
\usepackage{exscale}
\usepackage{amsbsy}
\usepackage{subfigure}
\usepackage{textcomp}
\usepackage{comment}
\usepackage{hyperref}
\usepackage{bm}
\usepackage{wrapfig}
\usepackage[font=footnotesize,labelfont=bf]{caption}
\usepackage{ulem} 
\usepackage{makecell}

\def\ba#1\ea{\begin{align}#1\end{align}}
\def\bg#1\eg{\begin{gather}#1\end{gather}}
\def\bm#1\em{\begin{multline}#1\end{multline}}
\def\bmd#1\emd{\begin{multlined}#1\end{multlined}}

\newcommand{\be}{\begin{equation}}
\newcommand{\ee}{\end{equation}}
\newcommand{\bea}{\begin{eqnarray}}
\newcommand{\eea}{\end{eqnarray}}

\newcommand{\pd}{\partial}

\newcommand{\matleft}{\left(\begin{array}}
\newcommand{\matright}{\end{array}\right)}
\newcommand{\Tr}{\operatorname{Tr}}

\newcommand{\sgn}{\operatorname{sgn}}

\usepackage[numbers,sort&compress]{natbib}
\def\simge{
    \mathrel{\rlap{\raise 0.511ex 
        \hbox{$>$}}{\lower 0.511ex \hbox{$\sim$}}}}

\def\simle{
    \mathrel{\rlap{\raise 0.511ex 
        \hbox{$<$}}{\lower 0.511ex \hbox{$\sim$}}}}

\makeatletter
\renewcommand\section{\@startsection {section}{1}{\z@}%
                                 {-3.5ex \@plus -1ex \@minus -.2ex}
                                   {2.3ex \@plus.2ex}%
                                   {\normalfont\large\bfseries}}
\renewcommand\subsection{\@startsection{subsection}{2}{\z@}%
                                   {-3.25ex\@plus -1ex \@minus -.2ex}%
                                     {1.5ex \@plus .2ex}%
                                     {\normalfont\bfseries}}
\renewcommand\subsubsection{\@startsection{subsubsection}{3}{\z@}%
                                   {-3.25ex\@plus -1ex \@minus -.2ex}%
                                     {1.5ex \@plus .2ex}%
                                     {\normalfont\itshape}}
\makeatother

\def\pplogo{\vbox{\kern-\headheight\kern -29pt
\halign{##&##\hfil\cr&{\ppnumber}\cr\rule{0pt}{2.5ex}&\ppdate\cr}}}
\makeatletter
\def\ps@firstpage{\ps@empty \def\@oddhead{\hss\pplogo}%
  \let\@evenhead\@oddhead 
}
\def\maketitle{\par
 \begingroup
 \def\thefootnote{\fnsymbol{footnote}}
 \def\@makefnmark{\hbox{$^{\@thefnmark}$\hss}}
 \if@twocolumn
 \twocolumn[\@maketitle]
 \else \newpage
 \global\@topnum\z@ \@maketitle \fi\thispagestyle{firstpage}\@thanks
 \endgroup
 \setcounter{footnote}{0}
 \let\maketitle\relax
 \let\@maketitle\relax
 \gdef\@thanks{}\gdef\@author{}\gdef\@title{}\let\thanks\relax}
\makeatother

\hypersetup{
    unicode=false,          
    pdftoolbar=true,        
    pdfmenubar=true,        
    pdffitwindow=false,     
    pdfstartview={FitH},    
    pdftitle={Jain in the LLL},    
    pdfauthor={},     
    pdfsubject={Subject},   
    pdfcreator={},   
    pdfproducer={}, 
    pdfkeywords={keyword1} {key2} {key3}, 
    pdfnewwindow=true,      
    colorlinks=true,       
    linkcolor=OliveGreen, 
    citecolor=NavyBlue,        
    filecolor=magneta,      
    urlcolor=cyan           
}

\numberwithin{equation}{section}

\newcommand*\samethanks[1][\value{footnote}]{\footnotemark}

\usepackage[latin1]{inputenc}

\usepackage{amsmath}
\usepackage{color}
\usepackage{amsfonts}
\usepackage{amssymb}
\usepackage[mathscr]{euscript}
\usepackage{graphicx}
\usepackage{geometry}
\usepackage{amssymb,epsfig,subfigure}
\usepackage{hyperref}
\usepackage{comment}
\usepackage{tabularx}
\usepackage{bm}
\usepackage{euscript}
\usepackage{graphicx}
\usepackage[dvipsnames]{xcolor}
\usepackage{amsfonts}
\usepackage{exscale}
\usepackage{amsbsy}
\usepackage{subfigure}
\usepackage{textcomp}
\usepackage{comment}
\usepackage{hyperref}
\usepackage{slashed}
\usepackage{authblk}
\usepackage{tabularx}
\usepackage{euscript}
\usepackage{graphicx}
\usepackage{color}
\usepackage{amsfonts}
\usepackage{exscale}
\usepackage{amsbsy}
\usepackage{subfigure}
\usepackage{textcomp}
\usepackage{comment}
\usepackage{hyperref}
\usepackage{bm}
\usepackage{wrapfig}
\usepackage[font=footnotesize,labelfont=bf]{caption}
\usepackage{ulem} 
\usepackage{makecell}
\def\ba#1\ea{\begin{align}#1\end{align}}
\def\bg#1\eg{\begin{gather}#1\end{gather}}
\def\bm#1\em{\begin{multline}#1\end{multline}}
\def\bmd#1\emd{\begin{multlined}#1\end{multlined}}

\usepackage[numbers,sort&compress]{natbib}
\def\simge{
    \mathrel{\rlap{\raise 0.511ex 
        \hbox{$>$}}{\lower 0.511ex \hbox{$\sim$}}}}

\def\simle{
    \mathrel{\rlap{\raise 0.511ex 
        \hbox{$<$}}{\lower 0.511ex \hbox{$\sim$}}}}

\makeatletter
\renewcommand\section{\@startsection {section}{1}{\z@}%
                                 {-3.5ex \@plus -1ex \@minus -.2ex}
                                   {2.3ex \@plus.2ex}%
                                   {\normalfont\large\bfseries}}
\renewcommand\subsection{\@startsection{subsection}{2}{\z@}%
                                   {-3.25ex\@plus -1ex \@minus -.2ex}%
                                     {1.5ex \@plus .2ex}%
                                     {\normalfont\bfseries}}
\renewcommand\subsubsection{\@startsection{subsubsection}{3}{\z@}%
                                   {-3.25ex\@plus -1ex \@minus -.2ex}%
                                     {1.5ex \@plus .2ex}%
                                     {\normalfont\itshape}}
\makeatother

\def\pplogo{\vbox{\kern-\headheight\kern -29pt
\halign{##&##\hfil\cr&{\ppnumber}\cr\rule{0pt}{2.5ex}&\ppdate\cr}}}
\makeatletter
\def\ps@firstpage{\ps@empty \def\@oddhead{\hss\pplogo}%
  \let\@evenhead\@oddhead 
}
\def\maketitle{\par
 \begingroup
 \def\thefootnote{\fnsymbol{footnote}}
 \def\@makefnmark{\hbox{$^{\@thefnmark}$\hss}}
 \if@twocolumn
 \twocolumn[\@maketitle]
 \else \newpage
 \global\@topnum\z@ \@maketitle \fi\thispagestyle{firstpage}\@thanks
 \endgroup
 \setcounter{footnote}{0}
 \let\maketitle\relax
 \let\@maketitle\relax
 \gdef\@thanks{}\gdef\@author{}\gdef\@title{}\let\thanks\relax}
\makeatother

\hypersetup{
    unicode=false,          
    pdftoolbar=true,        
    pdfmenubar=true,        
    pdffitwindow=false,     
    pdfstartview={FitH},    
    pdftitle={Jain in the LLL},    
    pdfauthor={},     
    pdfsubject={Subject},   
    pdfcreator={},   
    pdfproducer={}, 
    pdfkeywords={keyword1} {key2} {key3}, 
    pdfnewwindow=true,      
    colorlinks=true,       
    linkcolor=OliveGreen, 
    citecolor=NavyBlue,        
    filecolor=magneta,      
    urlcolor=cyan           
}

\numberwithin{equation}{section}

 \newcommand\beal{\begin{equation}\begin{aligned}}
\newcommand\eeal{\end{aligned}\end{equation}}

\begin{document}

\normalem

\setcounter{page}0
\def\ppnumber{\vbox{\baselineskip14pt
}}

\def\ppdate{
} \date{\today}

\title{\Large \bf   Lowest Landau level theory of the bosonic Jain states}
\author{Hart Goldman and T. Senthil}
\affil{\it \small Department of Physics, Massachusetts Institute of Technology, Cambridge, MA 02139, USA}

\maketitle

\begin{abstract}
Quantum Hall systems offer the most familiar 
setting where strong inter-particle interactions combine with the topology of single particle states to yield novel phenomena. Despite our mature understanding of these systems, an open challenge  has been to to develop a microscopic theory 
capturing both their universal and non-universal 
properties, 
when the Hamiltonian is restricted to the non-commutative space of the lowest Landau level. 
Here we develop such a theory for the Jain sequence of bosonic fractional quantum Hall states at fillings 
  $\nu={p\over p+1}$. 
  Building on a lowest Landau level description of a parent composite fermi liquid at $\nu = 1$, 
  we describe how to dope the system to reach the Jain states. Upon doping, 
  the composite fermions fill non-commutative generalizations of Landau levels, and the Jain states correspond to integer composite fermion filling.  Using this approach, we obtain an approximate  expression for the bosonic Jain sequence gaps with no reference to any long-wavelength approximation. Furthermore, we show that the universal properties, such as Hall conductivity, are encoded in an effective non-commutative Chern-Simons theory, which is obtained on integrating out the composite fermions. This theory has the same topological content as the familiar Abelian Chern-Simons theory on commutative space. 
\end{abstract}

\pagebreak
{
\hypersetup{linkcolor=black}
\tableofcontents
}
\pagebreak

\section{Introduction}
Two dimensional  many-particle systems in a strong magnetic field display some of the most famous examples of quantum correlated phenomena. At special, partial fillings $\nu$ of a Landau level, an incompressible phase is formed with Hall conductivity, $\sigma_{xy}=\nu e^2/h$, in what is known as the fractional quantum Hall effect (FQHE). 
Since Laughlin's original explanation of the basic physics of the FQHE \cite{Laughlin1983}, a number of distinct theoretical approaches have been developed which 
provide a deeper and more versatile understanding of the FQHE. 
A prominent and successful approach is built on the framework of flux attachment \cite{Wilczek-1982}, 
which trades the original theory of charged particles in a magnetic field for a theory of new entities, dubbed composite bosons or composite fermions, interacting with a Chern-Simons gauge field. In particular, the composite fermion construction provides a simple unifying 
explanation of the vast majority of observed FQH phases as integer quantum Hall (IQH) states of composite fermions \cite{Jain-1989, Lopez-1991, Jainbook}. 
It also provides the foundation to understand the metallic states found in electronic systems near even denominator filling fractions \cite{Halperin-1993, Halperin1996, Simon1998, Son2015, Goldman2018a, Halperin2020}, and enables parent descriptions of non-Abelian quantum Hall states \cite{Read2000,Goldman2020}. 

For the gapped quantum Hall states, the universal long distance properties are captured by Chern-Simons topological quantum field theories, enriched by a global $U(1)$ symmetry that corresponds to particle number conservation \cite{Xiaogangbook,Fradkinbook}.   Within the flux attachment framework, these Chern-Simons theories can be usefully viewed as arising from integrating out Landau levels of composite fermions. This description encapsulates the topological order and symmetry fractionalization data of the quantum Hall state. However, it is not suitable if we are interested in estimating non-universal, microscopic 
quantities, such as the magnitude of the gaps, given a microscopic Hamiltonian. 
Especially salient in this regard are Hamiltonians defined by projecting to the lowest Landau level (LLL). 
Then the electron kinetic energy is quenched and all energy scales are determined by the interaction strength. The interactions must therefore be treated completely non-perturbatively,  
and the problem is typically only amenable to study through a variety of numerical methods, such as variational wavefunction calculations, exact diagonalization, and the density matrix renormalization group. An interesting analytical approach, due to Murthy and Shankar 
(for a review, see Ref. \cite{Murthy2003}), 
used a Hamiltonian description of composite fermions in the LLL 
to yield good estimates of gaps and other non-universal quantities for a variety of filling fractions. Nevertheless, the universal 
properties of 
both the composite Fermi liquid 
and the Jain states seem harder to directly extract from this formalism. 

An important open question in quantum Hall physics is to obtain a unified analytic framework that captures both universal and non-universal aspects of the physics. This task is particularly 
challenging when the many-particle Hilbert space is restricted to a single Landau level. 
Recently, inspired by the earlier efforts of Pasquier, Haldane, and Read \cite{Pasquier1998, Read1998}, one of us, together with Zhihuan Dong, made progress on this problem by developing a composite fermion approach to the metallic state of bosons at $\nu=1$ that is strictly in the LLL \cite{Dong2020, Dong2021}. This theory, like earlier   approaches, involves composite fermions coupled to a fluctuating $U(1)$ gauge field, although this time the fields live on the non-commutative space of the LLL. 
Despite this progress for the metallic state, there continues to be no construction of gapped quantum Hall states that allows for a derivation of the long wavelength Chern-Simons field theory but also provides estimates of the gaps and other non-universal features. 

\begin{figure}
\includegraphics[width=\textwidth]{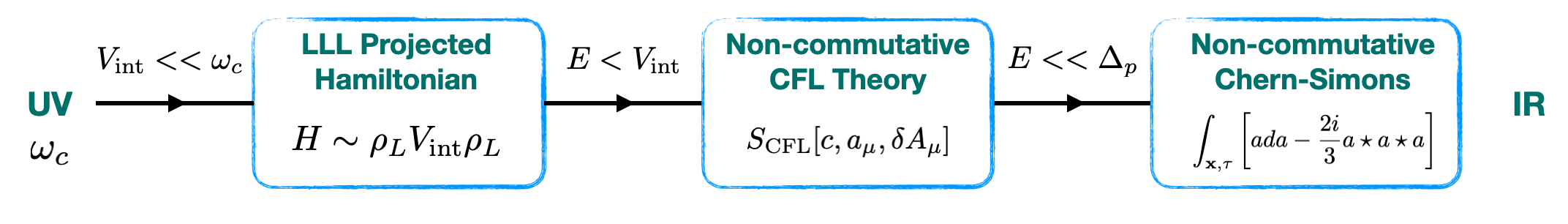}
\caption{A schematic of the emergence of the bosonic Jain states. Our starting point is a theory of bosons in a magnetic field with interaction energy, $V_{\mathrm{int}}$, and cyclotron frequency, $\omega_c$. The bosons are projected to the lowest Landau level (LLL) when $V_{\mathrm{int}}<<\omega_c$ (we use units $\hbar=c=1$). By changing variables to composite fermions, $c$, coupled to a fluctuating gauge field, $a_\mu$, we consider a composite Fermi liquid (CFL) theory on the non-commutative space of the LLL, which is valid below the interaction scale. We access the Jain states by turning on a background vector potential, $\delta A_i$, such that the composite fermions fill $p$ Landau levels with a gap, $\Delta_p$. Integrating out the composite fermions leads to an effective \emph{non-commutative} Chern-Simons theory, which captures the topological data of the Jain state.}
\label{figure}
\end{figure}

The purpose of this 
work is to provide such a theory for the specific case of Jain states of bosons at fillings $\nu = \frac{p}{p+1}$,  with $p$ a positive integer, which approach the metallic state of bosons at $\nu=1$ as $p\rightarrow\infty$. An outline of our construction is given in Figure \ref{figure}. Starting with the description of the metallic composite Fermi liquid ground state of bosons at $\nu = 1$ in terms of  non-commutative composite fermion field theory, we show that the proximate Jain states can be obtained by filling Landau levels of composite fermions, in a manner directly analogous to the derivation using ordinary composite fermions. 
In the process, we obtain an approximate  mean field expression for the bosonic Jain sequence gaps, 
\begin{align}
\Delta_p&= 
\frac{2\pi\bar\rho}{m_\star}\frac{1}{1+p}\,,
\end{align}
where $\bar\rho$ is the charge density, $m_*$ is the composite fermion effective mass, and $p=2\pi\bar\rho/\delta B$ is the Jain sequence index. This expression is corrected compared to the usual composite fermion mean field result, in which the gap is proportional to the effective magnetic field felt by the composite fermions, $m_*\Delta=\delta B=2\pi\bar\rho/p$. In our result, working on the non-commutative space of the LLL has corrected $p\rightarrow p+1$. Using the Hartree-Fock approximation of the effective mass, our result returns values for the gap that are compatible with exact diagonalization estimates \cite{Regnault2003, Regnault2004}. In principle, other non-universal data, such as collective modes and structure factors, can also be sought using our mean field framework. 

Furthermore, we show that integrating out the composite fermions yields a non-commutative Chern-Simons (NCCS) field theory for these Jain states, which describes the correct topological order. 
Many years ago, based on hydrodynamic considerations, Susskind and Polychronakos \cite{Susskind2001, Polychronakos2001} proposed the NCCS field theory and its matrix model regularization  to describe quantum Hall states at short distances,   
and it has been found to correctly capture ground state topological features and wave functions \cite{Hellerman2001,Karabali2001a,Karabali2002}. More recently, the quantum Hall matrix models have been generalized to the Jain sequences \cite{Cappelli2006} and to non-Abelian quantum Hall states  \cite{Dorey2016,Dorey2016a}. Their universal geometric response has also been studied \cite{Lapa2018, Lapa2018a}. However, the connection of the NCCS theory to realistic microscopic Hamiltonians and to composite fermion field theory has been opaque. These relationships are explained in our approach. The NCCS theory appears as an effective description at scales $\omega<<\Delta_p$, and 
its non-commutativity is set by the charge density, rather than the total magnetic field, in agreement with the hydrodynamic approach of Susskind \cite{Susskind2001}.

Since the universal properties of fractional quantum Hall phases are well described by the usual commutatice Chern-Simons field theory, one may 
question what is gained by using the non-commutative version. It is therefore interesting that in our work, which is an approximate treatment of microscopic Hamiltonians in the LLL, the non-commutative theory emerges rather naturally, arising as a bridge between a fully microscopic theory and the usual long wavelength 
topological quantum field theory.


We finally comment that although there is an extensive literature on non-commutative field theories (for reviews, see Refs. \cite{Douglas2001,Szabo2001}), the structure of the 
theories we encounter differs from what is most commonly discussed. 
In particular, our focus is on non-commutative theories of fermions 
at a non-zero density of global $U(1)$ charge and in a background magnetic field. Our work contributes to the general understanding of this class of problems. Nevertheless, we will make contact with and use some of the results of the extensive existing literature, as appropriate, throughout the paper.

We proceed as follows. In Section \ref{sec: review}, we review the non-commutative composite fermion approach the problem of bosons at $\nu=1$ engineered in Refs. \cite{Dong2020}. We then proceed to develop our mean field approach to the bosonic Jain sequence in Section \ref{sec: MFT}, in the process obtaining our result for the gaps. In Section \ref{sec: Jain}, we integrate out the composite fermions to obtain a NCCS field theory describing the universal aspects of the Jain sequence states. We conclude in Section \ref{sec: discussion}.

\section{Recap: The LLL composite fermion theory of bosons at $\nu=1$} 
\label{sec: review}

\subsection{Basics of the LLL and mean field theory of the metallic state at $\nu=1$}

We start by considering a 2d system of (bosonic or fermionic) charged particles with density $\bar\rho$ in a strong magnetic field, $B$, such that the filling, $\nu=2\pi\bar\rho/B$, is less than or equal to one. The particles then form highly degenerate Landau levels, and they can each fill states in the lowest Landau level. In the limit of large magnetic field, we may therefore project the Hilbert space to states in the LLL. On restricting to the LLL, the particles are only described by their guiding center coordinates, 
\begin{align}
R_i&=x_i-\ell_B^2\varepsilon_{ij}\pi_j\,, 
\end{align}
where $\ell_B=|B|^{-1/2}$ is the magnetic length and $\pi_i$ is the gauge invariant momentum. Assuming canonical commutation relations, these coordinates do not commute,
\begin{align}
\label{eq: non-commutative space}
[R_i,R_j]&=-i\ell_B^2\varepsilon_{ij}\equiv i\Theta\varepsilon_{ij}\,.
\end{align}
Therefore, the geometry of the LLL is non-commutative, with non-commutativity parameter $\Theta=-\ell_B^2$. 

On LLL projection, the kinetic energy is quenched, and the Hamiltonian may be expressed in terms of interactions of the projected density operator\footnote{The projection of the density operator will also include a form factor $e^{- q^2 l_B^2/4}$. We will follow the common practice of incorporating this form factor into the interaction so that the Hamiltonian is expressed in terms of the operator $\rho_L$ which satisfies the GMP algebra.} , which we denote $\rho_L(\mathbf{q})=\sum_n e^{i\mathbf{q}\cdot\mathbf{R}_n}$, where boldface denotes spatial vectors and $n$ is a particle index. Using Eq. \eqref{eq: non-commutative space}, the density operator can be seen to satisfy the famous Girvin-MacDonald-Platzman (GMP) algebra \cite{Girvin1986},
\begin{align}
\label{eq: GMP}
[\rho_L(\mathbf{q}),\rho_L(\mathbf{p})]&=2i\sin\left(\frac{\ell_B^2(\mathbf{q}\times\mathbf{p})}{2}\right)\rho_L(\mathbf{q}+\mathbf{p})\,.
\end{align}
The GMP algebra contains information on the geometric symmetries of the LLL: $\rho_L$ can be expanded in terms of the generators of area-preserving diffeomorphisms of space, which in turn satisfy the $\mathcal{W}_{\infty}$ algebra \cite{Cappelli1992, Iso1992, Haldane2009}. 
While these are not symmetries of the microscopic Hamiltonian they encode the structure of the non-commutative space of the LLL. 
Our interest is in projected Hamiltonians of the form, 
\begin{align}
\label{eq: microscopic H}
H&=\frac{1}{2} \int\frac{d^2\mathbf{q}}{(2\pi)^2}\tilde{U}(\mathbf{q})\rho_L(-\mathbf{q})\rho_L(\mathbf{q})\,.
\end{align}
 where $\tilde{U}(\mathbf{q}) = U_0 \exp{- \frac{q^2 l_B^2}{2}}$ is the Fourier transform of a contact interaction augmented with a form factor. 
We first review the solution of this problem developed in Ref. \cite{Dong2020} for the particular case of bosons at $\nu=1$.

The physical underpinnings of the lowest Landau level description is a view of the composite fermions as charge-vortex dipoles \cite{Read1994}. 
Unlike in ordinary flux attachment, these composite fermions are understood to be neutral, because the vortices deplete a unit of charge at their cores. They possess a dipole moment proportional to their momentum,
\begin{align}
\label{eq: dipole moment}
\mathbf{d}=\ell_B^2\mathbf{k}\times\hat{z}
\end{align}
The composite fermions thus naturally acquire a quadratic dispersion from their dipole energy,  $\epsilon(\mathbf{k})\sim|\mathbf{d}|^2$. We will refer back to this basic physical picture at many points in this work.

We begin with a formal  representation  of the many body bosonic Hilbert space and the density operator introduced by Pasquier and Haldane \cite{Pasquier1998}, and developed further by Read \cite{Read1998}.  For $N$ particles, any state in the many-body Hilbert space of of bosons at $\nu=1$ can be written as
\begin{align}
|\Psi\rangle=\sum_{\{m_i\}}a_{m_1\dots m_N}|m_1,\dots,m_N\rangle\,,
\end{align}
where $m_i=1,\dots,N$ is an orbital index for the $N$ single particle states in the Landau level (thus the filling $\nu=1$), $|m_1,\dots,m_N\rangle$ is a product of single particle orbitals, and $a_{m_1\dots m_N}$ are constants that are symmetric under exchange of indices. One may represent the basis states in terms of composite fermion creation (annihilation) operators, $c^\dagger_{mn}$ ($c_{mn}$), where $m$ and $n$ are again orbital indices,
\begin{align}
|m_1,\dots m_N\rangle&=\varepsilon^{n_1\dots n_N}c^\dagger_{m_1n_1}\dots c^\dagger_{m_Nn_N}|0\rangle\,,
\end{align}
where $|0\rangle$ is the vacuum state annihilated by $c_{mn}$. We may interpret the left and right indices, respectively, as being associated with (bosonic) charges and vortices, which bind to form the composite fermion. In terms of the fermion operators, the projected density operator is
\begin{align}
(\rho_L)_{mm'}=c^\dagger_{mn}c_{m'n}
\end{align}
and is the generator of the physical electromagnetic symmetry.

This fermionic description is redundant. The physical states are invariant under $SU(N)$ rotations of the $n_i$ indices, meaning that the theory has a $SU(N)$ gauge symmetry. The presence of this gauge symmetry implies a constraint, in that the physical states satisfy
\begin{align}
\label{eq: matrix constraint}
(\rho_R)_{nn'}|\Psi\rangle=c^\dagger_{mn}c_{mn'}|\Psi\rangle=\delta_{nn'}|\Psi\rangle\,.
\end{align}
Here $\rho_R$ is the $SU(N)$ charge density, and we may interpret it as the density of vortices in the LLL. Note that its diagonal part is shared with that of the charge density, $\rho_L$, and represents the physical EM charge, $\Tr[\rho_L]=\Tr[\rho_R]=N$. 

We may construct momentum-space representations of the fermion operators, $c(\mathbf{q})$, $c^\dagger(\mathbf{q})$ using the orbital matrix elements of the magnetic translation operator, $\tau_\mathbf{q}=e^{i\mathbf{q}\cdot\mathbf{R}}$. These operators satisfy the standard anticommutation relations
\begin{align}
\label{eq: anticommutation relations}
\{c(\mathbf{q}),c^\dagger(\mathbf{q}')\}=(2\pi)^2\delta^{(2)}(\mathbf{q}-\mathbf{q}')\,.
\end{align}
In terms of these operators, we may construct $\rho_L$ and $\rho_R$ as follows,
\begin{align}
\label{eq: q densities}
\rho_L(\mathbf{q})=\int\frac{d^2\mathbf{k}}{(2\pi)^2}c^\dagger(\mathbf{k}-\mathbf{q})c(\mathbf{q})e^{-i\Theta\mathbf{k}\times\mathbf{q}/2}\,&,\,\rho_R(\mathbf{q})=\int\frac{d^2\mathbf{k}}{(2\pi)^2}c^\dagger(\mathbf{k}-\mathbf{q})c(\mathbf{q})e^{i\Theta\mathbf{k}\times\mathbf{q}/2}\,.
\end{align}
where we recall $\Theta=-\ell_B^2=-1/(2\pi\bar\rho)$ at $\nu=1$. In this representation, the constraint in Eq. \eqref{eq: matrix constraint} becomes
\begin{align}
\label{eq: q constraint}
\rho_R(\mathbf{q})&=(2\pi)^2\bar\rho\,\delta^{(2)}(\mathbf{q})\,.
\end{align}
Using the anticommutation relation in Eq. \eqref{eq: anticommutation relations}, it is immediate that both of these operators furnish representations of the GMP algebra: $\rho_L$ satisfies Eq. \eqref{eq: GMP} while $\rho_R$ satisfies Eq. \eqref{eq: GMP} but with a minus sign on the right hand side. Using the constraint in Eq. \eqref{eq: q constraint} and expanding $\rho_L$ in powers of $\mathbf{q}$, it is straightforward to see that the composite fermions possess the dipole moment\footnote{Each term in the expansion of $\rho_L$ corresponds to a generator of diffeomorphisms in the LLL, which satisfy the $\mathcal{W}_\infty$ algebra. The dependence of the dipole moment on momentum is natural from this point of view: it is the generator of translations.} in Eq. \eqref{eq: dipole moment},
\begin{align}
d_i=\ell_B^2\,\varepsilon_{ij}\mathcal{P}^j\,,
\end{align}
where $\mathcal{P}$ is the composite fermion momentum operator.

Equipped with the fermionic representation of the density operator, we may revisit the Hamiltonian in Eq. \eqref{eq: microscopic H}, which is now a four-fermion interaction plus the constraint \eqref{eq: q constraint}. In Refs. \cite{Read1998,Dong2020,Dong2021} this problem was attacked using a Hartree-Fock mean field approach, in which the authors sought a ground state of fermions at finite density, consistent with the expectation of a composite Fermi liquid at $\nu=1$,
\begin{align}
\langle c^\dagger(\mathbf{q})c(\mathbf{q})\rangle&\neq 0\,.
\end{align}
This leads to a Hartree-Fock Hamiltonian of the form
\begin{align}
H_{\mathrm{HF}}&=\int\frac{d^2\mathbf{q}}{(2\pi)^2}\epsilon(\mathbf{q})c^\dagger(\mathbf{q})c(\mathbf{q})\,,
\end{align}
which describes a metallic state of composite fermions, with effective mass of order the interaction strength. The quadratic dispersion, $\epsilon(\mathbf{q})\sim|\mathbf{q}|^2$ at small $\mathbf{q}$, is naturally understood as the dipole energy of the composite fermion.

The assumption of a non-vanishing expectation value for the Hartree-Fock order parameter, $ c^\dagger(\mathbf{q})c(\mathbf{q})$, spontaneously breaks the $SU(N)$ gauge symmetry: it only commutes with $\rho_R(\mathbf{q})$ at $\mathbf{q}=0$. The important fluctuations about the Hartree-Fock ground state are therefore those near $\mathbf{q}=0$, which can be described by coupling to a $U(1)$ gauge field, $a_\mu$. However, this is not an ordinary $U(1)$ gauge symmetry, since the fields under consideration live on the non-commutative space of the LLL. Therefore, one must consider a non-commutative composite fermion field theory \cite{Dong2020,Dong2021}. This theory will anchor our analysis through the rest of this work, and we are now prepared to introduce it.


\subsection{Non-commutative composite fermion field theory and the Seiberg-Witten map}
To build a non-commutative field theory, it is necessary to understand how to multiply functions on non-commutative space. These requires the introduction of the Moyal star product. For functions of non-commutative coordinates, $x_i$, the Moyal star product is
\begin{align}
f(x)\star g(x)&=\lim_{x'\rightarrow x}\exp\left(i\frac{\Theta}{2}\varepsilon^{ij}\frac{\partial}{\partial x'^{\,i}}\frac{\partial}{\partial x^j}\right)f(x')g(x)\,.
\end{align}
Forming products in this way, the effective composite fermion action\cite{Dong2020,Dong2021}, including gauge field fluctuations, is
\begin{align}
\label{eq: CFL action}
S&=\int d\tau d^2\mathbf{x}\left[c^\dagger\star D_\tau c+\frac{1}{2m_*}(D_ic)^\dagger\star(D^ic)+ia_0\bar\rho\right]\,,\\
D_\mu c&=\pd_\mu c-i\,\delta A_\mu\star c-ic\star a_\mu\,.
\end{align}
Here 
$\delta A_\mu$ is the background \emph{probe} electromagnetic vector potential, such as, for instance,  that needed to described the deviation of the theory from $\nu=1$. On the other hand, $a_\mu$ is an emergent, fluctuating $U(1)$ gauge field. The equation of motion of $a_\tau$ enforces the constraint,
\begin{align}
\label{eq: constraint}
c^\dagger\star c=\bar\rho=\frac{B}{2\pi}\,,
\end{align}
where $B$ is the magnetic field felt by the underlying bosonic charges, which are at filling $\nu=1$ when $\delta A = 0$. The theory in Eq. \eqref{eq: CFL action} therefore describes a metallic state of composite fermions at density set by the background magnetic field.
 
Unlike in ordinary commutative field theory, on non-commutative space even $U(1)$ gauge symmetries have non-Abelian representations. The physical electromagnetic $U(1)$ symmetry acts to the left on the non-commutative composite fermions, 
\begin{align}  
\label{eq: U(1)L 1}
c&\rightarrow U_L\star c\,, \\
\label{eq: U(1)L 2}
\delta A&\rightarrow U_L\star \delta A\star U_L^\dagger-i\,\partial U_L\star U_L^\dagger\,,\\
\label{eq: U(1)L 3}
a &\rightarrow a\,,
\end{align}
while the emergent $U(1)$ gauge symmetry associated with the constraint acts to the right,
\begin{align}
\label{eq: U(1)R 1}
c&\rightarrow c\star U_R\,, \\
\delta A&\rightarrow \delta A\,,\\
\label{eq: U(1)R 3}
a &\rightarrow U_R^\dagger\star a\star U_R-i\,U_R^\dagger\star \partial U_R\,.
\end{align}
Respectively, these gauge symmetries correspond to the conserved charge densities,
\begin{align}
\rho_L=-c\star c^\dagger&,\rho_R=c^\dagger\star c\,,
\end{align}
On quantization, each of these density operators can be seen to satisfy their own GMP algebra, Eq. \eqref{eq: GMP}, as they correspond to the right and left densities introduced in the previous subsection.

Non-commutative field theories like Eq. \eqref{eq: CFL action} are exceptionally challenging to work with, especially in the presence of a constraint like Eq. \eqref{eq: constraint}. A common approach to studying them is to map them to a corresponding field theory on commutative space, using a mapping developed by Seiberg and Witten \cite{Seiberg1999}. This map relates  the fields $c,a,\delta A$ in Eq. \eqref{eq: CFL action} to new fields, $\psi,\hat{a},\delta\hat{A}$, defined on commutative space\footnote{Frequently in the literature on non-commutative field theory, the hat notation actually denotes the \emph{non-commutative} fields. For consistency with Refs. \cite{Dong2020, Dong2021}, we continue to use the inverted notation.}, in an  expansion in powers of the non-commutative parameter, $\Theta=-\ell_B^2$.
Remarkably, applying the Seiberg-Witten map 
to the non-commutative CFL theory in Eq. \eqref{eq: CFL action}, one obtains\cite{Dong2020} the Halperin-Lee-Read (HLR) theory \cite{Halperin-1993} for bosons at $\nu=1$, but with additional short-ranged corrections,
\be
\label{eq: SW HLR}
\hat{\mathcal{L}}=\psi^\dagger(\partial_\tau-i\hat{a}_\tau-i\delta \hat{A}_\tau)\psi+\frac{1}{2m_*}|(\pd_i-i\hat{a}_i-i\delta\hat{A}_i)\psi|^2-i\frac{1}{4\pi}\hat{a}d\hat{a}+\Theta\mathcal{L}_{\mathrm{corr}}\,.
\ee
Via the Seiberg-Witten map, the constraint term, $ia_\tau\bar{\rho}$, in the non-commutative CFL theory becomes a Chern-Simons term in the commutative theory, since $2\pi\bar\rho\Theta=-\nu=-1$. Flux attachment in the HLR theory can therefore be thought of as ``emerging'' from the constraint on $\rho_R$, Eq. \eqref{eq: constraint}, in the non-commutative CFL theory! The terms in $\Theta\mathcal{L}_{\mathrm{corr}}$ are fixed by the Seiberg-Witten map but are sub-leading in powers of $\Theta$, meaning that they are irrelevant at long wavelengths. For more details on this calculation, see Ref. \cite{Dong2020}.

The Seiberg-Witten map thus exchanges a theory of composite fermions on non-commutative space, whose density has a non-trivial form factor, for a theory of ordinary composite fermions augmented by additional short-ranged terms. This theory is therefore easily amenable to doping away from $\nu=1$, resulting in the usual Jain sequence of bosonic FQH states, in for which the composite fermions fill an integer number of Landau levels,
\begin{align}
\nu=\frac{p}{p+1}\,.
\end{align}
Because this theory corresponds to a non-commutative CFL theory in the LLL, one can also calculate estimates for LLL dynamical features at large $p$ (weak $\delta \hat{B}\equiv\delta \hat{F}_{xy}$). For example, using this theory, one can calculate the gaps for the Jain sequence states. One finds\footnote{We note that Ref. \cite{Dong2020} contains a sign error in its final expression for the gaps. We correct it here.}
\begin{align}
\label{eq: SW gaps}
\Delta(\delta \hat{B})&=\frac{|\delta \hat{B}|}{m_*}\left(1+\Theta\delta\hat{B}\right)+\mathcal{O}\left((\Theta \delta\hat{B})^2\right)=\frac{2\pi\bar\rho}{m_*}\left(\frac{1}{p}-\frac{1}{p^2}\right)+\mathcal{O}(p^{-3})\,,
\end{align}
where we have used the fact that in the Jain state the filling fraction of the composite fermions is $\nu_{\mathrm{CF}}=2\pi\bar\rho/\delta\hat{B}=p$. In this work we will go much further by doping the non-commutative CFL theory, Eq. \eqref{eq: CFL action}, directly. This will enable us to compute dynamical properties of the LLL bosonic Jain states, such as their gaps, in a mean field approximation valid to all orders in $\Theta$. We further show that our results match the long wavelength estimate of Eq. \eqref{eq: SW gaps}. In addition, we will obtain the low energy effective theory for gauge fluctuations about the mean field. This will enable us to provide a microscopic derivation of the non-commutative Chern-Simons theory description of quantum Hall states and make contact with prior work that proposed such a description based on hydrodynamic arguments \cite{Susskind2001,Polychronakos2001}.

\section{Doping away from $\nu=1$}
\label{sec: MFT}

\subsection{A physical picture: quantum mechanics of charge-vortex dipoles}
\label{sec: single particle}

Before turning on a non-commutative magnetic field in the composite fermion field theory of Eq. \eqref{eq: CFL action}, a clear physical picture can be formed by considering the single particle problem of charges and vortices in the LLL, with the composite fermions viewed as charge-vortex dipoles. The Jain states are then obtained by filling Landau levels of the composite fermions. 
Remarkably, we will show later on that this framework corresponds precisely to the single particle limit of the non-commutative composite fermion field theory. 

Technically, our analysis of the charge-vortex dipole problem will rely on the classic work of Nair and Polychronakos \cite{Nair2001} which solved the Landau level problem for particles moving in a non-commutative space.  This work has  since seen considerable followup, e.g. in Refs. \cite{Karabali2001,Morariu2001,Correa2001,Christiansen2001,Kokado2002,Chakraborty2003,Kokado2004,Delduc2007}. Although physical quantum Hall systems and composite fermion theory are occasionally mentioned in this literature, the precise connection to this physics has not been made before as far as we know. We explain this connection below.

Consider bosonic charges and vortices in the $\nu=1$ system, respectively characterized by guiding center coordinates, $R_i$ and $R^v_i$. Because the system has a finite charge density, the vortices see the same magnetic field but with opposite sign to the charges. Hence their coordinates satisfy an algebra with opposite sign,
\begin{align}
\label{eq: doped guiding center algebra}
[R_i,R_j]=-i\ell_B^2\,\varepsilon_{ij}\,,\,[R^v_i,R_j^v]=+i\ell_B^2\,\varepsilon_{ij}\,,[R_i,R^v_j]=0\,.
\end{align}
We also define ``composite fermion" coordinates, $r^{\mathrm{CF}}_i=(R_i+R_i^v)/2$, which commute,
\begin{align}
[r^{\mathrm{CF}}_i,r^{\mathrm{CF}}_j]&=0\,.
\end{align}
As expected, the composite fermion feels no magnetic field. Since the composite fermion is a dipole of a charge and a vortex, we may define a  dipole moment, $d_i=R^v_i-R_i$, which is proportional to the canonical momentum of the composite fermion,
\begin{align}
\label{eq: undoped CF momentum}
p^{\mathrm{CF}}_i=\frac{1}{\ell_B^2}\varepsilon_{ij}d^j\,,\,[r^{\mathrm{CF}}_i,p^{\mathrm{CF}}_j]=i\delta_{ij}\,,\,[p^{\mathrm{CF}}_i,p^{\mathrm{CF}}_j]=0\,.
\end{align}
It is thus natural to take the effective Hamiltonian to be the dipole energy,
\begin{align}
\label{eq: single particle H}
H&=\frac{1}{2m_*}(p^{\mathrm{CF}})^2\,.
\end{align}
Indeed, this expectation is borne out in the Hartree-Fock analysis of Ref. \cite{Dong2020}.

We now investigate how this picture changes on varying the background magnetic field, $B\rightarrow B+\delta B$, while keeping the charge density fixed so that $\nu_{\mathrm{CF}}=2\pi\frac{\bar\rho}{\delta B}=\frac{B}{\delta B}=p$, where $p$ is an integer. The filling fraction of the bosonic charges is therefore
\begin{align}
\nu=2\pi\frac{\bar\rho}{B+\delta B}=\frac{p}{1+p}\,,
\end{align}
which is the bosonic Jain sequence. Tuning the magnetic field alters the guiding center algebra of the charges while preserving that of the vortices,
\begin{align}
[R_i,R_j]=-i\ell_{B+\delta B}^2\,\varepsilon_{ij}\,,\,[R^v_i,R_j^v]=+i\ell_B^2\,\varepsilon_{ij}\,,[R_i,R^v_j]=0\,,
\end{align}
where $\ell_{B+\delta B}^2=1/(B+\delta B)$. Now the composite fermions see a magnetic field, and their coordinates no longer commute,
\begin{align}
[r^{\mathrm{CF}}_i,r^{\mathrm{CF}}_j]&=i\frac{1}{4}(\ell_B^2-\ell_{B+\delta B}^2)\,\varepsilon_{ij}\equiv i\theta\,\varepsilon_{ij}\,.
\end{align}
The composite fermion momentum defined in Eq. \eqref{eq: undoped CF momentum} now  satisfies
\begin{align}
[r^{\mathrm{CF}}_i,p^{\mathrm{CF}}_j]=\frac{\ell_{B+\delta B}^2+\ell_B^2}{2\ell_B^2}\,i\delta_{ij}\,,\,[p^{\mathrm{CF}}_i,p^{\mathrm{CF}}_j]=i\frac{\ell_B^2-\ell_{B+\delta B}^2}{\ell_B^4}\varepsilon_{ij}=i\frac{\delta B}{1+\delta B/B}\varepsilon_{ij}\,.
\end{align}
 Notice that the momentum commutator no longer vanishes, corresponding to the fact that the composite fermions are experiencing a magnetic field. This commutation algebra, together with the Hamiltonian in Eq. \eqref{eq: single particle H}, is precisely the non-commutative Landau level problem studied by Nair and Polychronakos \cite{Nair2000} (after a trivial rescaling  of the momentum, which we will perform below, 
 to make it canonically conjugate to the composite fermion coordinate). 

As in the ordinary Landau problem, we may diagonalize the Hamiltonian in Eq. \eqref{eq: single particle H} by constructing creation and annihilation operators out of the momentum, proportional to $p_x^{\mathrm{CF}}\pm i p_y^{\mathrm{CF}}$. This allows one to easily compute the gaps, which are 
\begin{align}
\Delta_p &=\frac{\delta B}{m_*}\frac{1}{1+B/\delta B}=\frac{2\pi\bar\rho}{m_*}\frac{1}{1+p}\,.
\end{align}
Amazingly, we will find that analysis and the resulting gaps precisely match that of the non-commutative composite fermion mean field theory we will develop in the next subsection.

Implicit in the above analysis is the identification of $\delta B$ with the variation of the physical magnetic field. Hence we expect the density of states of the non-commutative Landau level to be $\delta B$. This can be checked by constructing an operator, $\hat{\mathcal{D}}_i$, which commutes with the Hamiltonian and generates magnetic translations of $r^{\mathrm{CF}}$, the composite fermion guiding center coordinate \cite{Nair2000}. To do this, we start by defining a new momentum that satisfies canonical commutation relations with $r^{\mathrm{CF}}$,
\begin{align}
\tilde{p}^{\mathrm{CF}}&=\frac{2\ell_B^2}{\ell^2_{B+\delta B}+\ell_B^2}\,p^{\mathrm{CF}}\,,
\end{align}
Using $\tilde{p}^{\mathrm{CF}}$, the theory is now defined by the algebra,
\begin{align}
[r^{\mathrm{CF}}_i,r^{\mathrm{CF}}_j]=i\theta\,\varepsilon_{ij}\,,[r^{\mathrm{CF}}_i,\tilde{p}^{\mathrm{CF}}_j]=i\delta_{ij}\,,\,[\tilde{p}^{\mathrm{CF}}_i,\tilde{p}^{\mathrm{CF}}_j]=4i\frac{(\ell_B^2-\ell_{B+\delta B}^2)}{(\ell^2_{B+\delta B}+\ell_B^2)^2}\varepsilon_{ij}\equiv i\delta B_{\mathrm{CF}}\,\varepsilon_{ij}\,,
\end{align}
It is now easy to see that the operators, $\mathcal{D}_i=(\tilde{p}_i^{\mathrm{CF}}-\delta B_{\mathrm{CF}}\varepsilon_{ij}r^{\mathrm{CF}}_j)/(1-\theta\delta B_{\mathrm{CF}})$, commute with the Hamiltonian. These operators satisfy the algebra,
\begin{align}
[\mathcal{D}_x,\mathcal{D}_y]=-i\frac{\delta B_{\mathrm{CF}}}{1-\theta\delta B_{\mathrm{CF}}}=-i\delta B\,.
\end{align}
This is a strong indication that the density of states of the Landau level is $\delta B$. 
Further discussion of the degeneracy of Landau levels on the non-commutative torus can be found in 
Refs. \cite{Connes1997,Ho1998,Morariu1998,Morariu2001}, and we sketch the derivation in Appendix \ref{appendix: degeneracy}. 
Furthermore, we will observe below that the change in the physical magnetic field here, $\delta B$, can be identified with the  the magnetic field obtained by mapping a non-commutative background field to an ordinary one via the Seiberg-Witten map.

Using this single particle picture of charge-vortex dipoles to physically ground us, we will now 
derive these same results using the non-commutative composite fermion theory in Eq. \eqref{eq: CFL action}, which we dope using a non-commutative vector potential. In doing so, we will demonstrate how the introduction of a uniform non-commutative magnetic field in the field theory picture corresponds to the deformation of the non-commutativity of the charge degrees of freedom presented above.

\subsection{Mean field theory of the composite fermion Landau problem}
\subsubsection{Mean field Hamiltonian and Landau level spectrum}
We now construct the bosonic Jain sequence by doping the non-commutative field theory, Eq. \eqref{eq: CFL action} and studying the resulting problem in mean field theory. While it is not possible to alter the density of the composite fermions given the constraint, Eq. \eqref{eq: constraint}, one can still turn on a background magnetic field associated with $\delta A_i$. Indeed, if we neglect fluctuations of $a_i$, the non-commutative composite fermion theory in a uniform magnetic field is remarkably similar to the analogous problem of  electrons on an ordinary space. The composite fermions continue to form Landau levels, with gaps set by the cyclotron frequency. Furthering the analogy, when small fluctuations of $a_i$ are introduced, we will show in Section \ref{sec: Jain} that integrating out the filled Landau levels leads to a \emph{non-commutative} Chern-Simons theory. This will ultimately lead us to the bosonic Jain sequence states.


We start by passing to a Hamiltonian formulation, fixing the density via the constraint, Eq. \eqref{eq: constraint}. Neglecting fluctuations of $a_i$, i.e. $\langle a_i\rangle=0$, the mean field Hamiltonian can be written as 
\begin{align}
\label{eq: mean field H}
\mathcal{H}&=\int d^2\mathbf{x}\,\frac{1}{2m_*}(D_ic)^\dagger\star D^ic\,,D_ic=\partial_ic-i\delta A_i\star c\,.
\end{align}
The equation of motion of this theory leads to the single particle Schr\"{o}dinger equation, 
\begin{align}
\label{eq: effective LL problem}
Hc=-\frac{1}{2m_*}D_iD^ic=Ec\,.
\end{align}
Rather than fixing to a specific gauge and computing the spectrum by solving the resulting differential equation, it is more conceptually straightforward in this non-commutative context to solve the theory simply by considering its algebraic properties, as in Ref. \cite{Nair2000} and the analysis of the previous subsection. The commutator of the $D_i$ operators is the non-commutative field strength,
\begin{align}
\delta F_{xy}=i[D_x,D_y]_\star=\partial_x(\delta A_y)-\partial_y(\delta A_x)-i[A_x,A_y]_\star\,,
\end{align}
where we have defined the star commutator, $[A,B]_\star=A\star B-B\star A$. To solve the Schr\"{o}dinger equation, Eq. \eqref{eq: effective LL problem}, for uniform $\delta F_{xy}$, we can again construct creation and annihilation operators using $D_x\pm i D_y$, leading to Landau levels with energy set by $\delta F_{xy}$,
\begin{align}
\label{eq: LL energies}
\epsilon_n=\omega_c\left(n+\frac{1}{2}\right)\,,
\end{align}
where
\begin{align}
\label{eq: MFT cyclotron}
\omega_c(\delta F_{xy}) 
=\frac{\delta F_{xy}}{m_*}
\end{align}
is the Landau level gap. This expression superficially differs from the result for Landau level gaps in Eq. \eqref{eq: SW gaps}. However, we stress that the non-commutative field strength, $\delta F_{xy}$, is not equal to the physical magnetic field, which sets composite fermion Landau level degeneracy and corresponds to the physical magnetic field, $\delta B$, in Section \ref{sec: single particle}. We now describe how to relate these two quantities. 

\subsubsection{Some formalism: Covariant coordinates}

To define the filling fraction for the composite fermion Landau levels, it is necessary to determine their degeneracy. While na\"{i}vely 
one may expect that the degeneracy is given by $\delta F_{xy}\times\mathrm{Area}$, this is in fact not the case. Indeed, in the analysis of the dipole model in Section \ref{sec: single particle}, the field setting the energy gap was found to be differ from the physical magnetic field. A hint of this can be seen in the fact that the commutator of the ordinary coordinate operators, $x_i$, with the gauge covariant momenta, $D_i$, is not gauge invariant: $D_i$ transforms as an adjoint while $x_i$ does not transform. Non-commutative gauge invariance therefore dictates that we work with coordinate operators that transform under gauge transformations. These are the so-called covariant coordinates, 
\begin{align}
\label{eq: Y covariant coordinate definition}
Y^i[\delta A]&=x^i+\Theta\,\varepsilon^{ij}\delta A_j\,,
\end{align}
which transform as adjoints under the left-acting gauge transformations, Eq. \eqref{eq: U(1)L 2},
\begin{align}
\label{eq: Y covariant coordinate transformation law}
Y^i\rightarrow U_L\star Y\star U_L^\dagger\,,
\end{align}
ensuring gauge invariant commutation relations with $D_i$. 

The transformation law follows from the close relationship between non-commutative gauge transformations and area-preserving diffeomorphisms (APDs) of the non-commutative space.  
For a more detailed review of this topic, see Ref. \cite{Douglas2001}. Consider a gauge transformation, $U_L=e^{i\lambda}$. Then for any function, $f(\mathbf{x})$,
\begin{align}
\label{eq: APD}
U_L\star f(\mathbf{x})\star U_L^\dagger&=f(\mathbf{x})-\Theta\,\varepsilon^{ij}\partial_i\lambda\,\partial_jf(\mathbf{x})+\mathcal{O}(\Theta^2)=f(x^i+\Theta\varepsilon^{ij}\partial_j\lambda+\dots)\,.
\end{align}
This means that a combination of left and right gauge transformations is infinitesimally equivalent to a translation by a vector, $\Theta\,\varepsilon^{ij}\partial_j\lambda$, which leaves the area element invariant. Therefore, under a left-acting gauge transformation, the covariant coordinate, $Y^i$, transforms as an adjoint,
\begin{align}
Y^i\rightarrow x^i+\Theta\,\varepsilon^{ij} U_L\star \delta A_j\star U_L^\dagger+\Theta\,\varepsilon^{ij}\,\partial_j\lambda+\mathcal{O}(\Theta^2)=U_L\star Y^i\star U_L^\dagger\,.
\end{align}
A useful consequence of this transformation law is that functions of covariant coordinates also transform as adjoints,
\begin{align}
\label{eq: f[Y] transformation}
f[Y(\mathbf{x})]\rightarrow f\left[U_L(\mathbf{x})\star Y(\mathbf{x})\star U_L^\dagger(\mathbf{x})\right]=U_L(\mathbf{x})\star f[Y(\mathbf{x})]\star U_L^\dagger(\mathbf{x})\,.
\end{align}
This property follows immediately from the equivalence of adjoint gauge transformations with APDs, Eq. \eqref{eq: APD}, and it will figure heavily in Section \ref{sec: Jain}. 

\subsubsection{Bosonic Jain sequence and gaps}

We are now prepared to extract the Landau level degeneracy and physical magnetic field, as well as make contact with the dipole quantum mechanics problem in Section \ref{sec: single particle}. 
We start by replacing operators that act with a Moyal product to the right with operators defined to act with left multiplication, such that all operators only act to the left on states \cite{Nair2000}. 
We may then define new coordinate operators, which we (suggestively) name $R^i$ and $R^v_i$, such that
\begin{align}
R_ic=Y_i\star c\,&,\,R^v_ic=c\star x_i\,.
\end{align}
These operators furnish two different, mutually commuting guiding center algebras, which may be naturally expressed in terms of a new magnetic field,
\begin{align}
\label{eq: SW B}
\delta\hat{B}&=\frac{\delta F_{xy}}{1+\Theta\delta F_{xy}}\,,
\end{align}
such that
\begin{align}
[R_i,R_j]=i\Theta(1+\Theta\delta F_{xy})\varepsilon_{ij}&=-\frac{i}{B+\delta \hat{B}}\varepsilon_{ij}=-i\ell_{B+\delta\hat{B}}^2\varepsilon_{ij}\,,\\
[R^v_i,R^v_j]=i\ell_B^2\,\varepsilon_{ij} &,
[R_i,R^v_j]=0\,, \nonumber
\end{align}
where we recall $\Theta=-\ell_B^2$. This is the same guiding center algebra as in the dipole quantum mechanics model, Eq. \eqref{eq: doped guiding center algebra}, with $\delta \hat{B}$ identified with the physical magnetic field, which we had earlier denoted $\delta B$.

The mean field Hamiltonian, Eq. \eqref{eq: effective LL problem}, can also be expressed in terms of $R_i$ and $R_i^v$, since in non-commutative field theory commutators with $x^i$ are derivatives,
\begin{align}
[x^i,c(\mathbf{x})]_\star=i\Theta\,\varepsilon^{ij}\partial_jc(\mathbf{x})\,.
\end{align}
Therefore, the covariant derivative can be written as
\begin{align}
\label{eq: covariant derivative coord rep}
D^ic=\frac{i}{\Theta}\,\varepsilon^{ij}(R_j-R^v_j)c\,,
\end{align}
meaning that it is proportional to the composite fermion dipole moment, as anticipated in Section \ref{sec: single particle}. In mean field theory, the single particle Hamiltonian of the composite fermions is therefore identical to the Hamiltonian in Eq. \eqref{eq: single particle H},
\begin{align}
\label{eq: mapped dipole H}
H=\frac{1}{2m_*}\frac{1}{\ell_B^4}(R_i-R^v_i)^2\,,
\end{align}
The mean field composite fermion theory and the charge-vortex dipole problem are equivalent! We can therefore immediately apply the result \cite{Nair2000} for the Landau level degeneracy in Section \ref{sec: single particle} to find that it is indeed set by $\delta \hat{B}$,
\begin{align}
\label{eq: degeneracy}
d_{LL}
=\frac{\delta \hat{B}}{2\pi}\times \mathrm{Area}\,.
\end{align}
See 
Refs. \cite{Connes1997,Ho1998,Morariu1998,Morariu2001} and Appendix \ref{appendix: degeneracy} for a more detailed account of how one arrives at this result for the case of the theory on a torus. We note that the field strength, $\delta\hat{B}$, is also the value of the field strength obtained from the Seiberg-Witten map, meaning that this result is consistent with the requirement of flux quantization in the commutative approximation of the theory \cite{Dirac1931, Wu1976}.  

To summarize, by using the proper gauge covariant coordinates, we have shown that the introduction of the non-commutative vector potential in the composite fermion mean field theory corresponds precisely to the deformation of the non-commutative parameter of the charges in the single-dipole model, with Eq. \eqref{eq: SW B} as the physical magnetic field. With this result, we can define the composite fermion filling fraction,
\begin{align}
\nu_{\mathrm{CF}}&=2\pi \frac{\langle \rho_L\rangle}{\delta \hat{B}}=2\pi\frac{\bar\rho}{\delta \hat{B}}\,.
\end{align}
The bosonic Jain states occur when the composite fermions form integer quantum Hall states, with $\nu_{\mathrm{CF}}=p$, $p$ an integer. 

Because $\delta\hat{B}$ sets the degeneracy of the composite fermion Landau level, we interpret it as shift in the magnetic field felt by the underlying bosonic charges from $B=2\pi\bar\rho$. Therefore, one is naturally led to the conclusion that the filling fraction of the physical bosons is
\begin{align}
\label{eq: bosonic Jain fillings from mft}
\nu=2\pi\frac{\bar\rho}{B+\delta \hat{B}}=\frac{p}{1+p}\,.
\end{align}
This is precisely the expected form of the bosonic Jain sequence \cite{Regnault2003,Regnault2004}! 
In terms of the density $\bar\rho$ and composite fermion Landau level index $p$, one finds that the gap for each Jain sequence state is thus,
\begin{align}
\label{eq: Jain gaps MFT}
\Delta_p= 
\frac{2\pi\bar\rho}{m_\star}\frac{1}{1+p}\approx\frac{2\pi\bar\rho}{m_*}\left(\frac{1}{p}-\frac{1}{p^2}\right)+\mathcal{O}(p^{-3})\,.
\end{align}
This expression, for which we have not invoked any long wavelength approximation, is one of the main results of this work. It matches the result from the Seiberg-Witten map, Eq. \eqref{eq: SW gaps}, but involves no long wavelength approximation and is thus valid even at small values of $p$.

Notably, for the $\nu=1/2$ Laughlin state, we find
\begin{align}
\Delta_{1}&=\frac{\pi\bar\rho}{m_\star}\approx(0.65)\,\pi \,U_0\bar\rho\,. 
\end{align}
This result uses the Hartree-Fock effective mass from Ref. \cite{Dong2020} for the case of a local contact potential and relies on mean field theory, $\langle a\rangle=0$, but nevertheless incorporates the non-commutativity of the LLL exactly. It is comparable to the value obtained with exact diagonalization, which is $[0.095(5)]\,2\pi U_0\bar\rho/\nu\approx (0.38)\pi U_0\bar\rho$. \cite{Regnault2003,Regnault2004}.  We note, however, that the effective mass can in principle depend on the magnetic field, $m_*(\delta \hat{B})$, which will lead to corrections to our result, particularly for states far away from the compressible state (small $p$) and very close to the compressible state (large $p$), where gauge fluctuations likely play an important role. In Section \ref{sec: discussion}, we will comment further on the physics leading to such a magnetic field dependent mass and how they could be handled theoretically to improve our result.

We now briefly comment on the states with negative composite fermion filling, $p<0$, which  correspond to $\nu > 1$. For $p=-1$, which in HLR theory corresponds to a superfluid, $\delta F_{xy}$ diverges, and so too does the gap in Eq. \eqref{eq: bosonic Jain fillings from mft}. This indicates a singularity of the non-commutative field theory, which presupposes $|\Theta|$ as the minimum uncertainty of the $x$ and $y$ coordinates \cite{Nair2000,Polychronakos2002}. In other words, all field strengths are cut off by the non-commutativity of space at scales of $1/|\Theta|$. The non-commutative field strength, $|\delta F_{xy}|$, continues to exceed $1/|\Theta|$ until $p=-2$, which in HLR corresponds to the $\nu=2$ bosonic integer quantum Hall state \cite{senthil2013integer}, and so the non-commutative composite fermion field theory is problematic for $-2<\nu_{\mathrm{CF}}\leq0$. 

Before concluding this section, we note that it would have been quite challenging to directly confirm that the magnetic field felt by the underlying bosonic charges is indeed $B+\delta \hat{B}$ using the Pasquier-Haldane-Read formalism, as the composite fermion operators become rectangular matrices when the filling deviates from $\nu=1$. However, the correspondence with the simple charge-vortex dipole model makes it clear that this is the only valid option.  
Moreover, we will demonstrate that the topological orders associated with the states at filling $\nu_{\mathrm{CF}}=p$ correspond precisely to that of the $p^{\mathrm{th}}$ bosonic Jain state. This is the topic we now turn to.

\section{The bosonic Jain sequence: Universal features}
\label{sec: Jain}


\subsection{Fluctuations and Hall response}
\label{sec: linear response}

Having concluded that doping the non-commutative composite fermion field theory, Eq. \eqref{eq: CFL action}, leads to the bosonic Jain sequence of fractional quantum Hall states, we are now prepared to assess their universal properties, which for ordinary composite fermions are encoded in a Chern-Simons effective field theory for the gauge fluctuations. In the original work of Lopez and Fradkin \cite{Lopez-1991} (based on the usual flux attachment transformation to composite fermions without restricting to the LLL), the Chern-Simons effective action was directly calculated by integrating out the composite fermions and expanding the resulting functional determinant. We follow the same logic for the composite fermions obtained in the present LLL construction. In this section, we present  a physically transparent derivation, in which we consider linear response starting from the first quantized dipole Hamiltonian, Eq. \eqref{eq: mapped dipole H}, and match the result to an effective \emph{non-commutative} Chern-Simons theory. In Appendix \ref{appendix: determinant}, we present a more formal derivation by calculating the polarization tensor that determines the quadratic part of the effective action for the gauge fields when the composite fermions are integrated out.   We perform our calculations  without fixing to a particular gauge, so non-commutative gauge invariance is manifest throughout.

We begin by introducing fluctuations into the Hamiltonian in Eq. \eqref{eq: mapped dipole H}. This amounts to replacing the vortex guiding center coordinates, $R^v_i$, with their covariant counterparts,
\begin{align}
R^v_ic=c\star X_i[a]\,,
\end{align}
where
\begin{align}
X^i[a]=x^i-\Theta\,\varepsilon^{ij} a_j
\end{align}
are the covariant coordinates for the fluctuating gauge field. Under a right-acting gauge transformation, Eq. \eqref{eq: U(1)R 3}, they transform as 
\begin{align}
X^i\rightarrow U_R^\dagger\star X^i\star U_R\,,
\end{align}
and their commutator is
\begin{align}
[X_i,X_j]_\star=i\Theta(\varepsilon_{ij}-\Theta f_{ij})\,,
\end{align}
although we will be primarily interested in situations where $f_{ij}=\partial_i a_j-\partial_ja_i+i[a_i,a_j]_\star=0$. With this new definition for $R^v_i$, the covariant derivative continues to take the form of Eq. \eqref{eq: covariant derivative coord rep}. Therefore, the single particle Hamiltonian retains the form of Eq. \eqref{eq: mapped dipole H}, but now it is understood to transform in the adjoint representation under both left and right-acting gauge transformations. 

We now study the response of the charges and vortices to the physical and emergent electric fields, using the first quantized description. If $V_b[R]$ and $V_v[R^v]$ are the scalar potentials felt by the physical charges and the vortices respectively, the Hamiltonian is
\begin{align}
\label{eq: H with E fields}
H&=\frac{1}{2m_*}\frac{1}{\ell_B^4}(R_i-R_i^v)^2+V_b[R_i]+V_v[R^v_i]\,.
\end{align}
Because both $R_i$ and $R^v_i$ are defined in terms of the covariant coordinates, the scalar potentials  are each adjoints under gauge transformations, as in Eq. \eqref{eq: f[Y] transformation}. In terms of the field theory representation, we may therefore consider them as background values of $\delta A_0$ and $a_0$. We emphasize that the inclusion of scalar potentials for both the charges and vortices is essential, since we will see that in order to satisfy the constraints implemented by the gauge fluctuations, as in Eq. \eqref{eq: constraint}, the vortices and the charges will simultaneously exhibit a Hall effect. The same phenomenon occurs in ordinary flux attachment: within the FQH state, establishing an electric field for the physical charges leads to an average electric field for the emergent (statistical) gauge field.

The introduction of the scalar potentials leads to currents of boson charges and vortices. In the Heisenberg picture, the world-line of a boson (vortex) is $R_i(t)$ ($R^v_i(t)$), so we define the current densities,
\begin{align}
\label{eq: first quantized currents}
j^b_i=\bar\rho\,\left\langle\frac{d R_i}{dt}\right\rangle&,\,\,j^v_i=\bar\rho\,\left\langle\frac{d R^v_i}{dt}\right\rangle\,.
\end{align}
In particular, we consider $j_i^b$ (which is an adjoint under left-acting gauge transformations), to be the physical current density. We also define the physical and emergent electric fields as
\begin{align}
E^i&=-\frac{i}{\ell_{B+\delta \hat{B}}^2}\,\left\langle\varepsilon^{ij}\left[R_j,V_b[R]\right]\right\rangle\equiv-\left\langle \partial_{R_i}V_b[R]\right\rangle\,,\\
e^i&=\frac{i}{\ell_B^2}\left\langle\varepsilon^{ij}\left[R_j^v,V_v[R^v]\right]\right\rangle\equiv-\left\langle\partial_{R_i^v}V_v[R^v]\right\rangle\,,
\end{align}
For constant magnetic fields (as in the case of interest), these are simply the gradients of the scalar potentials. One can confirm the physically intuitive conclusion that $E$ is the physical electric field by noticing that, if we write $V_b[Y]=\delta A_0(\mathbf{x})$, where $Y(\mathbf{x})$ is the covariant coordinate and $\delta A_0(\mathbf{x})$ is a static scalar potential, then we can use the star commutator with $Y$ to relate
\begin{align}
-\ell_{B+\delta \hat{B}}^2\,\partial_{Y_i} V_b[Y]=\ell_B^2\,\delta F_{0i}\,,
\end{align}
since $\delta A_i$ is time-independent. For a uniform electric field, this leads to 
\begin{align}
\delta F_{0i}=\frac{1}{1+\delta\hat{B}/B}E_i\,.
\end{align}
One can check using the formulas in e.g. Ref. \cite{Dong2020} that this is the relation between the non-commutative electric field, $\delta F_{0i}$, and the ordinary Abelian electric field of the Seiberg-Witten map. Since we have already commented on how the Seiberg-Witten gauge field satisfies proper flux quantization, this means that $E$ can be considered the physical electric field.   

We compute the DC response to 
uniform electric fields by solving the Heisenberg equations of motion. To make the equations more compact, we reintroduce the notation, $p^{\mathrm{CF}}_i=\frac{1}{\ell_B^2}\varepsilon^{ij}(R^v_j-R_j)$, and we drop terms which vanish for $\langle f_{ij}\rangle=0$. Then we obtain 
\begin{align}
\label{eq: drift}
\frac{d p^{\mathrm{CF}}_i}{dt}&= \frac{1}{i}\,[p^{\mathrm{CF}}_i,H]=\delta F_{xy}\,\varepsilon_{ij}\frac{p^{\mathrm{CF}}_j}{m_*}-\frac{1}{\ell_B^2}\left(\ell^2_{B+\delta \hat{B}}\,\partial_{R_i}V_b[R]+\ell_B^2\,\partial_{R^v_i}V_v[R^v]\right)\,,\\
\label{eq: boson velocity}
\frac{dR_i}{dt}&=\frac{1}{i}[R_i,H]=\frac{\ell_{B+\delta \hat{B}}^2}{\ell_B^2}\frac{p^{\mathrm{CF}}_i}{m_*}-\ell_{B+\delta \hat{B}}^2\,\varepsilon_{ij}\partial_{R_j}V_b[R]\,,\\
\label{eq: vortex velocity}
\frac{dR^v_i}{dt}&=\frac{1}{i}[R^v_i,H]=\frac{p_i^{\mathrm{CF}}}{m_*}+\ell_B^2\,\varepsilon_{ij}\partial_{R^v_i} V_v[R^v]\,.
\end{align}
The first equation determines the composite fermion drift velocity, $p^{\mathrm{CF}}/m_*$, while the latter two equations determine the individual charge and vortex responses. 

In addition to the equations of motion, the theory also has the constraint in Eq. \eqref{eq: constraint}, along with the equation of motion for $a_i$, which in the field theory sets $\langle J_R\rangle=\frac{i}{2m}\langle c^\dagger\star Dc-(Dc)^\dagger\star c\rangle=0$. Physically, we can understand these constraints  as the requirement that the vortices are fixed to have filling $\nu=-1$. In the Pasquier-Haldane-Read language, this is the requirement that the number of vortex orbitals is fixed to the number of physical bosons, even on tuning the physical filling away from $\nu=1$. The constraint can therefore be recast as the requirement that the vortices have unit Hall conductivity,
\begin{align}
j_v^i=-\frac{1}{2\pi}\varepsilon^{ij}e_j\,.
\end{align}
Plugging this back into Eq. \eqref{eq: vortex velocity}, we see that this is equivalent to the statement,
\begin{align}
\langle p^{\mathrm{CF}}\rangle&=0\,.
\end{align}
Because $p^{\mathrm{CF}}$ is proportional to the (gauge covariant) composite fermion dipole moment, the boson and the vortex coordinates sit on top of each other. With this constraint, Eq. \eqref{eq: drift} leads to a relation between the physical and emergent electric fields,
\begin{align}
\label{eq: E to e}
E_i=-\frac{\ell_B^2}{\ell_{B+\delta \hat{B}}^2}e_i=-\frac{p+1}{p}\,e_i\,.
\end{align} 
Looking to Eq. \eqref{eq: boson velocity}, we immediately obtain the physical Hall conductivity,
\begin{align}
j_b^i&=\frac{1}{2\pi}\bar\rho\,\ell_{B+\delta\hat{B}}^2\,\varepsilon^{ij}E_j=\frac{1}{2\pi}\frac{p}{p+1}\varepsilon^{ij}E_j\,.
\end{align}
Introducing units, the Hall conductivity is
\begin{align}
\label{eq: Hall conductivity}
\sigma_{xy}=\nu\,\frac{e^2}{h}=\frac{p}{p+1}\frac{e^2}{h}\,.
\end{align}
Hence the non-commutative composite fermion theory indeed leads to the correct Hall conductivity for the bosonic Jain sequence states!

\subsection{Non-commutative Chern-Simons theory}

Equipped with the result for the Hall conductivity in Eq. \eqref{eq: Hall conductivity} and the relation between the electric fields in Eq. \eqref{eq: E to e}, we can construct an effective Chern-Simons action at long wavelengths that reproduces them as the equations of motion,
\begin{align}
\label{eq: long wavelength CS}
S_{\mathrm{eff}}=-i\int d^2\mathbf{x}d\tau\,\left[\frac{p+1}{4\pi}ada+\frac{p}{2\pi}adA'+\frac{p}{4\pi}A'dA'+\mathcal{O}(\Theta)\right]\,,
\end{align}
where we use the notation $AdB=\varepsilon^{\mu\nu\lambda}A_\mu\partial_\nu B_\lambda$. To connect with the discussion above, $a_\mu$ is simply the fluctuating gauge field, while $A'=(A'_0(\mathbf{x}),0)$ is a probe field on top of the background field $\delta A$ that gives rise to the (physical) electric field, $E_i=-\partial_iA'_0$. The equation of motion for $a_i$ reproduces the relation in Eq. \eqref{eq: E to e}, and integrating out $a$ altogether returns the Hall response in Eq. \eqref{eq: Hall conductivity}.  

Importantly, at no point have we actually invoked the Seiberg-Witten map to an Abelian gauge theory. Indeed, the electric fields used in the analysis of Section \ref{sec: linear response} transform as adjoints under the \emph{non-commutative} gauge symmetries, Eqs. \eqref{eq: U(1)L 2} and \eqref{eq: U(1)R 3}. Therefore, the true effective action should display full non-commutative gauge invariance. Attempting to construct such an action leads to significant complications: because the left and right-acting gauge transformations are non-Abelian, there appears to be no gauge invariant mutual Chern-Simons term that may be represented in terms of star products of local operators. We will comment more on the pursuit of a non-commutative mutual Chern-Simons term in Section \ref{sec: BF}. 

For the purposes of diagnosing the topological order\footnote{Statements about topological order here have the caveat that $a$ couples to a fermion field in the fundamental representation. For ordinary commutative gauge theories, this can be formally captured by viewing  $a$  as a spin$_c$ connection rather than as an ordinary  $U(1)$ gauge field. For an explanation of this concept in a condensed matter context, see Refs. ~\cite{metlitski2015s,seiberg2016duality,senthil2019duality,Goldman2020}.  For the topological order, this means that the quasiparticle statistics are shifted by $\pi$ compared to the usual formulas (equivalently,  there is understood to be an additional spin-$1/2$ Wilson line).}, we may simply set $A'=0$ and construct a gauge invariant action for the fluctuating gauge field, $a$. 
The only such action that can be represented in terms of star products of local operators is the non-commutative Chern-Simons (NCCS) theory \cite{Susskind2001,Polychronakos2001},
\begin{align}
\label{eq: NCCS action}
S_{\mathrm{NCCS}}&=-\int d^2\mathbf{x}d\tau\,\frac{i(p+1)}{4\pi}\,\varepsilon^{\mu\nu\lambda}\left[a_\mu\star \partial_\nu a_\lambda+\frac{2i}{3}a_\mu\star a_\nu \star a_\lambda\right]\,.
\end{align}
As in non-Abelian gauge theories, gauge invariance dictates that this theory has a cubic interaction term even though the gauge group is $U(1)$. 
In fact, we can motivate the presence of the cubic term using the relation in Eq. \eqref{eq: E to e}.  If we turn off $E$, this equation becomes (dropping the brackets)
\begin{align}
0&=-\partial_{R^v_i}V_v[R^v]=\frac{i}{\ell_B^2}\varepsilon^{ij}\left[R_j^v,V_v[R^v]\right]\,.
\end{align}
Now if we identify $V_v[X]$ with a fluctuation of $a_0(\mathbf{x})$ and take $a_i(\mathbf{x})$ to be static, the commutator with the covariant coordinate, $X_i$, gives the field strength,
\begin{align}
0&=-\partial_i a_0+i[a_0,a_i]_\star=f_{0i}\,.
\end{align}
This matches the equation of motion of the NCCS theory, where the commutator originates from differentiating the cubic term.

The NCCS theory, particularly in its representation of a matrix model \cite{Susskind2001, Polychronakos2001}, has been extensively discussed as a short-wavelength description of fractional quantum Hall phases. 
However, the connection of these models to realistic microscopic Hamiltonians 
has been obscure. What is unique here is that we have obtained this theory as a \emph{long-wavelength} effective field theory of the bosonic Jain states that incorporates the non-commutativity of the lowest Landau level. Our result thus explains the connection between the non-commutative Chern-Simons theory and realistic microscopic models of quantum Hall phases, which until now was poorly understood. Furthermore, we have found that the non-commutativity of the Chern-Simons theory is set by the charge density, $\bar\rho$ (since $\Theta=-1/2\pi\bar\rho$), rather than the total magnetic field, $B+\delta \hat{B}$, in agreement with Susskind's original proposal \cite{Susskind2001}.

The topological ground state properties, such as anyons and their braiding, of NCCS theory are the same as the familiar Abelian Chern-Simons theory on commutative space. Indeed, it has been shown that at the classical level the NCCS action is equivalent to ordinary Chern-Simons action\footnote{The non-commutative theory actually contains gauge transformations that are singular in the corresponding commutative, Seiberg-Witten mapped theory, which lead to quantization of the level even on the plane   \cite{Nair2001, SheikhJabbari2001, Polychronakos2002}.} under the Seiberg-Witten map \cite{Grandi2000, Polychronakos2002} (a similar result was derived for the corresponding Wess-Zumino-Witten models \cite{Moreno2000}), and perturbative calculations have suggested that this correspondence extends to the quantum level as well \cite{Kaminsky2003}. Furthermore, the quantum Hall matrix models can be seen to reflect the correct topological order: for example, for the Laughlin states, Polychronakos demonstrated the existence of quasihole states with the correct fractional charge \cite{Polychronakos2001}. 

We note also that another argument for the emergence of NCCS theory was made in Ref. \cite{Fradkin2002}, which obtained a cubic interaction with Moyal phase factors using ordinary composite fermions (as in Ref.  \cite{Lopez-1991}) and proposed an emergent non-commutative gauge symmetry. 
Our conclusion contrasts with the result in Ref. \cite{Fradkin2002}, since the non-commutative gauge symmetry we consider is incorporated \emph{a priori} into the parent microscopic theory. Indeed, we do not believe there is any reason for non-commutative gauge symmetry to emerge unless it is inhereted from a short distance, LLL theory. 

There is also an interesting parallel between the first quantized composite fermion Hamiltonian we considered in Eq. \eqref{eq: H with E fields},  which is stated in terms of covariant world-line coordinates, and the quantum Hall matrix models. Indeed,  covariant worldline coordinates are also the basic variables in the matrix model description of non-commutative Chern-Simons theory, in which the definition of the covariant coordinates as non-commuting coordinates plus gauge fields is implemented dynamically \cite{Susskind2001, Polychronakos2001}. However, unlike the usual analysis of the matrix models, we did not introduce a regulator at long distances in order to convert the covariant coordinate operators to finite-dimensional matrices. It would be interesting in the future to explore what can be learned from applying such an approach to our composite fermion Hamiltonian.


Before moving on, we pause to make a technical comment regarding the specification of the topological order described by Eq. \eqref{eq: NCCS action}. In discussing ordinary non-Abelian Chern-Simons theory, one must be careful to specify a regularization. Due to the cubic interaction, the choice of regularization at short distances can 
lead to a one-loop exact shift in the non-Abelian Chern-Simons level \cite{Pisarski1985,Witten1988,Chen1992}, which is matched by an analogous quantum shift in in the corresponding Wess-Zumino-Witten model, where it appears in the computation of the central charge. The NCCS theory is no different. Perturbative calculations using a Maxwell regulator have found a level shit of $k\rightarrow k+\sgn(k)$ in $U(1)$ NCCS theory with level $k$ \cite{Chen2000}, although this does not affect the topological order \cite{AlvarezGaume2002}. The same shift\footnote{We note that Ref. \cite{SheikhJabbari2001} found the one-loop shift due to the Maxwell regulator to be $k\rightarrow k+2\sgn(k)$, based on some differences in normalization with Ref. \cite{Chen2000}. This difference can be properly settled by computing the free energy in the large-$k$ limit using the background field formalism, as in Ref. \cite{Witten1988}. We leave this for future work.} arises in the matrix models, where it comes from normal ordering a constraint \cite{Polychronakos2001}, and in this case it does affect the topological order (and shift the filling fraction accordingly). For our purposes, the NCCS level in Eq. \eqref{eq: NCCS action} is meant to reflect the full quantum Chern-Simons level, i.e. we implicitly choose a regulator in which no such shift appears.

\subsection{Toward a non-commutative mutual Chern-Simons theory}
\label{sec: BF}

We now revisit the question of how to construct an effective Chern-Simons action for both the background ($\delta A$) and fluctuating ($a$) gauge fields. As we commented above, because the left and right-handed non-commutative $U(1)$ gauge symmetries are not Abelian, there is no \emph{local} gauge invariant mutual Chern-Simons term. However, it should be possible to construct a non-local mutual Chern-Simons term, one which leverages the inherent non-locality of field theories on non-commutative space. Unfortunately, despite much effort, particularly in Refs. \cite{Chen2001,Hansson2003}, this term has proven elusive. While we will not completely solve this problem here, we will propose an action (which is not necessarily the generalization of Eq. \eqref{eq: long wavelength CS}) that is at least gauge invariant to $\mathcal{O}(\Theta)$. We expect that the intuition underlying this construction may prove useful in the pursuit of a full solution to this problem.

The basic problem with constructing a mutual Chern-Simons theory on non-commutative space is the same as in ordinary non-Abelian gauge theory: gauge invariance would necessitate that both participating gauge fields transform simultaneously, which does not appear possible by definition. However, with covariant coordinates, it is possible to \emph{induce} left-acting gauge transformations on the right-handed gauge field and \emph{vice versa}. For example, right-acting gauge transformations act as APDs on $\delta A[X]$, 
\begin{align}
\delta A[X]&\rightarrow \delta A\left[U(\mathbf{x})\star X\star U^\dagger(\mathbf{x})\right]=U(\mathbf{x})\star\delta A[X]\star U^\dagger(\mathbf{x})\,,
\end{align}
and analogously for $a[Y]$.

Leveraging this property, we can define a non-commutative Chern-Simons action that is gauge invariant to $\mathcal{O}(\Theta)$ under both left and right-acting gauge transformations by using the combination $a[Y]+\delta A[X]$\,,
\begin{align}
\label{eq: mutual CS proposal}
S_{\mathrm{NCCS}}\left[a[Y]+\delta A[X]\right]\,,
\end{align}
where $S_{\mathrm{NCCS}}$ is defined in Eq. \eqref{eq: NCCS action}. Notice that such an action is fully gauge invariant under right (left) gauge transformations if $Y$ ($X$) is replaced with the coordinate $\mathbf{x}$, at the cost of breaking invariance under the other gauge group. 

The reason the action in Eq. \eqref{eq: mutual CS proposal} is only invariant to $\mathcal{O}(\Theta)$ stems from the fact that it is non-local. Full gauge invariance under e.g. right-acting gauge transformations would require 
\begin{align}
\delta A[X]\rightarrow U^\dagger[Y]\star \delta A[X]\star U[Y]=U^\dagger(\mathbf{x})\star\delta A[X]\star U(\mathbf{x})+\mathcal{O}(\Theta^2)\,,
\end{align}
where the $\mathcal{O}(\Theta^2)$ term is non-vanishing. Resolving this issue would require introducing new Wilson line-like operators $V,W$ which transform under both left and right gauge transformations as follows
\begin{align}
U(1)_L&:\,V\rightarrow U^\dagger[X]\star V\star U(\mathbf{x})\,,\,W\rightarrow U[X]\star W\star U^\dagger[X]\,,\\
U(1)_R&:\, V\rightarrow U[Y]\star W\star U^\dagger[Y]\,,\, W\rightarrow U^\dagger[Y]\star W\star U(\mathbf{x})\,.
\end{align}
The gauge invariant Chern-Simons action would then be
\begin{align}
S_{\mathrm{NCCS}}\left[V^\dagger\star a[Y]\star V+W\star\delta A[X]\star W^\dagger\right]\,.
\end{align}
Unfortunately, we have not been able to construct explicit expressions for the operators, $V$ and $W$, and we leave this for future work. It is also not clear to us how to construct more general $K$-matrices than $(p,p,p)$ using this approach. We finally note that a discussion of non-commutative Chern-Simons theories with such $K$-matrices can be found in Ref. \cite{Chen2001}. However, the authors of that work fiat the mixed transformation laws for each gauge field, instead of attempting to induce them using covariant coordinates. The theories discussed in Ref. \cite{Chen2001} therefore cannot be obtained using the composite fermion approach outlined in our work, in which left and right gauge transformations do not mix.

\section{Discussion}
\label{sec: discussion}

A major challenge in quantum Hall physics 
has been to develop a microscopic theory that is defined in the lowest Landau level and is capable of 
capturing both universal and non-universal physics. In this work, we have  met this challenge for the specific case of bosonic Jain sequences at fillings $\nu=\frac{p}{p+1}$  using a composite fermion construction \cite{Pasquier1998,Read1998,Dong2020,Dong2021} that, unlike the standard flux attachment, explicitly lives in the  lowest Landau level. Previous work \cite{Read1998,Dong2020} employed this construction to discuss the metallic composite fermi liquid state for bosons at $\nu = 1$. An effective field theory description for this state 
consists of composite fermions coupled to a $U(1)$ gauge field on the non-commutative space of the LLL. 
Starting from this description, 
we doped the theory away from $\nu = 1$ to access the Jain sequence states. This is achieved by subjecting the composite fermions to a background, non-commutative magnetic field while holding their density fixed. 
Integrating out the composite fermions, we obtained a non-commutative Chern-Simons field theory, which encodes the topological features of the ground state. This conclusion significantly clarifies long-standing questions about the role of non-commutative Chern-Simons theory in the study of the fractional quantum Hall effect at short distances. It is a low-energy effective theory arising from integrating out composite fermions in the LLL. It captures the correct topological order of the ground state but does not contain any dynamical information on its own.  
Its non-commutativity is set by the charge density, as in the original proposal of Susskind \cite{Susskind2001}.

Our microscopic approach 
incorporates both the universal and non-universal data of quantum Hall states within a single theoretical framework. As an important demonstration, we presented an elegant, closed-form expression for the Jain sequence gaps, invoking only a mean field approximation.  It should also be possible to extract other dynamical features as well, such as the dispersion of the the GMP mode or the momentum dependence of the static structure factor. In approaching such calculations, it is important to note that in non-commutative field theory the density and current operators are not gauge invariant at finite momentum, meaning that some care will be necessary to ensure that results are gauge invariant. Another related  problem that will be important to attack in the future is the structure of the mutual Chern-Simons term in non-commutative field theory, which we expect to lack a representation in terms of star products of local operators. Another problem that could be tackled within our description is to study the evolution between the Jain ststes and a bosonic superfluid state by turning on a periodic potential in the LLL (along the lines of what was done at filling $\nu = 1$ in Ref. \cite{Dong2021}). 

The success of our approach for the bosonic Jain sequence invites the question of how to extend our framework to the fermionic Jain sequences. This would require a fully LLL theory of the composite Fermi liquid states at even denominator fillings in fermionic systems. Constructing such a theory is of great importance. For example, it would shed light on the emergence of a particle-hole symmetric composite Fermi liquid theory at $\nu=1/2$, like the Dirac theory proposed by Son \cite{Son2015} (an analogous ``reflection symmetry" was proposed by one of us for the states at $\nu=1/2n$ \cite{Goldman2018a}, but the status of that symmetry on LLL projection in clean systems is an open question; for alternate proposals in the LLL limit, see Ref. \cite{wang2016composite}). A non-commutative, LLL field theory of the $\nu=1/2$ state was proposed recently \cite{Milica2021}, but further work is needed in this direction.

We now comment on how our approximate result for the bosonic Jain sequence gaps in Eq. \eqref{eq: Jain gaps MFT} can be improved. 
In obtaining Eq. \eqref{eq: Jain gaps MFT}, we took the composite fermion effective mass, $m_*$, to be given by the result of the Hartree-Fock calculation at $\nu = 1$. A better approximation 
would be to calculate the effective mass directly at the filling of the Jain state, which would reveal if the effective mass has a field dependence, $m_*=m_*(\delta B)$. 
This could alter the dependence of the gaps, $\Delta_p$, on the Jain state index, $p$, from the form in Eq. \eqref{eq: Jain gaps MFT}. Such a calculation of $m_*(\delta B)$ can conceivably be 
performed within the Hamiltonian theory of Murthy and Shankar \cite{Murthy2003}. Indeed, we may regard the Hartree-Fock calculation within the 
 Hamiltonian theory as providing an improved mean field ansatz on top of which fluctuation effects can be included 
 using the non-commutative field theory. 
 
 Gauge fluctuations can also lead to field dependence of the effective mass. In the composite Fermi liquid itself, these 
 fluctuations lead to a diverging effective mass. As emphasized 
 by HLR in Ref. \cite{Halperin-1993}, on moving to proximate Jain fractions, this divergence will be cut-off at an energy scale given by the Jain gap, $\Delta_p$. This leads to a $\Delta_p$ that behaves (for small $\delta B$ and short ranged interactions)  as 
 $\Delta_p \sim |\delta B|^{\frac{3}{2}}\sim |p|^{-3/2}$. However, this asymptotic form is likely to only be relevant for very small $|\delta B|$, i.e. very large $p$. In the present problem of the bosonic Jain states, describing the region of large $p$ using the composite Fermi liquid will be additionally problematic, as the true ground state at $\nu = 1$ is the paired Pfaffian state. The large $p$ region will then involve competition between pairing and Landau level formation of composite fermions, and this will determine the details of the gap sizes and other non-universal characteristics. Therefore, for $p$ that is not too large, 
 we expect that the mean field description used in this work will 
 be adequate. 

Finally, a rich subject that we have not yet touched on is the response of quantum Hall systems to spatial curvature, which straddles universal and dynamical data \cite{Wen1992,Haldane2009,Haldane2011}. 
The Hall viscosity, or the parity-odd response to shears, is not universal \emph{a priori}, but is determined by the Wen-Zee shift -- a universal quantity -- in Galilean invariant quantum Hall states \cite{Wen1992,Read2009,Read2011}. Indeed, the matrix model regularization of non-commutative Chern-Simons theory has been shown to yield the correct Hall viscosity (up to an orbital contribution) for the Laughlin states and some non-Abelian quantum Hall states \cite{Lapa2018, Lapa2018a}. These arguments indicate that the non-commutative Chern-Simons theories we obtain at low energies will encode the Hall viscosity for the bosonic Jain states. This can also be checked by adapting the response calculations in Section \ref{sec: linear response} to finite wave vector, as the Hall viscosity appears in the coefficient of the $\mathcal{O}(q^2)$ contribution to the Hall conductivity \cite{Hoyos2011,Bradlyn2012}. 

A major open problem in this area is to obtain a microscopic derivation of the coupling of composite fermions to geometry, the so-called ``orbital spin" of the composite fermion.  A framework like ours that can bridge the gap between short and long-wavelength physics is an ideal platform on which to 
solve this problem. However, an obstacle to such a construction is inherent to non-commutative gauge theories: because of the relationship between gauge symmetry and area-preserving diffeomorphisms, it is generally not possible to construct a gauge invariant (or covariant) stress tensor that satisfies a local continuity equation \cite{AbouZeid2001} (a non-commutative equivalent to the Belinfante procedure is not known to us), and the construction of a Wen-Zee term suffers from similar challenges to the ordinary mutual Chern-Simons term. Resolving these questions and developing a non-commutative composite fermion field theory including a coupling to curvature will be an important direction for future studies.

\section*{Acknowledgements}

We thank Zhihuan Dong, Eduardo Fradkin, Prashant Kumar, and Hong Liu for discussions and comments on the manuscript. HG is supported by the Gordon and Betty Moore Foundation EPiQS Initiative through Grant No. GBMF8684 at the Massachusetts
Institute of Technology. This work was supported by NSF grant DMR-1911666,
and partially through a Simons Investigator Award from
the Simons Foundation to Senthil Todadri. This work was also partly supported by the Simons Collaboration on Ultra-Quantum Matter, which is a grant from the Simons Foundation (651440, TS).

\appendix

\section{Landau level degeneracy on the non-commutative torus}
\label{appendix: degeneracy}
In this Appendix, we sketch the derivation of the degeneracy of Landau levels on the non-commutative torus, which proceeds in analogy to the derivation on the ordinary torus. For more details, see Refs. \cite{Connes1997,Ho1998,Morariu1998,Morariu2001}, as well as the extensive literature on field theory on the non-commutative torus and $T$-duality, which is reviewed in Ref. \cite{Douglas2001}.

Without loss of generality, we choose to work on the square torus and identify,
\begin{align}
\label{eq: torus shift}
x_i\sim x_i+2\pi R\,,
\end{align}
where $R$ is the compactification radius. We define the physical position operators by exponentiation,
\begin{align}
V_i=e^{ix_i/R}\,,
\end{align}
which are invariant under shifts of $2\pi R$. They satisfy the algebra
\begin{align}
V_1V_2=V_2V_1e^{-i\Theta/R^2}\,.
\end{align}

We work with the single particle Hamiltonian in Eq. \eqref{eq: mean field H}, but we now choose to work in Landau gauge, 
\begin{align}
(\delta A_x,\delta A_y) &= (0,\delta F_{xy}x)\,,
\end{align}
The covariant derivatives, $D_i=\partial_i-i\delta A_i$, satisfy 
\begin{align}
[D_i,D_j]=-i\delta F_{xy}\,.
\end{align}
Because the covariant derivatives involve $x$ rather than $U_2=e^{ix/R}$, they transform under shifts of $x$ as
\begin{align}
\label{eq: D shift}
D_y\rightarrow D_y-i\delta F_{xy}(2\pi R)\,.
\end{align}
As in the case of the ordinary torus, this shift can be eliminated by a suitable gauge transformation, but now such gauge transformations are \emph{non-commutative}. The non-commutativity of the gauge group will ultimately be what alters the degeneracy from the non-commutative gauge flux.

The covariant derivative transforms as an adjoint under non-commutative (left-acting) gauge transformations. In Landau gauge, it changes as
\begin{align}
D_y\rightarrow U\star D_y\star U^\dagger=\partial_y+U\star \partial_y U^\dagger-i\delta F_{xy}\, U\star x\star U^\dagger\,.
\end{align}
To determine the gauge transformation that cancels the shift in Eq. \eqref{eq: D shift}, let $U=e^{-i\alpha y/R}$. Then
\begin{align}
D_y\rightarrow \partial_y-i\delta F_{xy}x+i\left(1+\Theta \delta F_{xy}\right)\frac{\alpha}{R}\,.
\end{align}
To cancel the shift in Eq. \eqref{eq: D shift}, we therefore must have
\begin{align}
\alpha = \frac{1}{2\pi}\frac{\delta F_{xy}}{1+\Theta \delta F_{xy}}\times (2\pi R)^2= \frac{\delta\hat{B}}{2\pi}\times (\mathrm{Area})\,.
\end{align}
$\alpha$ is therefore the flux of the ``physical" magnetic field defined in Eq. \eqref{eq: SW B} through the torus! Requiring that the gauge transformation $U$ itself be periodic on the torus therefore yields the flux quantization condition,
\begin{align}
\label{eq: torus flux quantum}
\oint_{T^2}\frac{\delta \hat{B}}{2\pi}=\frac{1}{2\pi}\frac{\delta F_{xy}}{1+\Theta\delta F_{xy}}\times (2\pi R)^2=n\in\mathbb{Z}\,.
\end{align}
Now, under the transformation $U$, the wave function on non-commutative space, $\Psi[x,y]$, transforms as $\Psi\rightarrow U\star \Psi$, so we seek a complete set of wave functions with the following properties,
\begin{align}
\Psi[x,y+2\pi R]&=\Psi[x,y]\\
\Psi[x+2\pi R,y]&=U^\dagger\star\Psi[x,y]\,.
\end{align}
The space of such wave functions constitutes the space of degenerate ground states on the torus. In ordinary, commutative space, the ground states correspond to the set of theta functions, and the dimension of the space of ground states (the LL degeneracy) is given by the flux piercing the torus. In this case as well, the space of ground states (the fundamental sections of the non-commutative gauge theory on the torus) has dimension set by the number of flux quanta, $|n|$ \cite{Connes1997,Ho1998,Morariu1998}. This completes the argument that the Landau level degeneracy is set by the magnetic field in Eq. \eqref{eq: SW B}. 

\section{Chern-Simons effective action}
\label{appendix: determinant}

In this Appendix, we show explicitly that integrating out $p$ non-commutative composite fermion Landau levels leads to an effective Chern-Simons action with level $p+1$ for the fluctuating gauge field, $a_\mu$, as in Eq. \eqref{eq: long wavelength CS}.  We do this by computing the polarization tensor, $\pi_{xy}(\omega)$, in the uniform ($\mathbf{q}\rightarrow0$) limit. The other components of the polarization tensor are then fixed by gauge invariance. Additionally, while we do not compute the cubic term in the non-commutative Chern-Simons action here (it is only nonzero at finite wave vector), it is also required to appear by gauge invariance.

For spatially uniform fluctuations, $a_i(\tau)$, the coupling to the composite fermions is
\begin{align}
\Gamma_{ac^\dagger c}&=\int d^2\mathbf{x}d\tau\,\frac{i}{2m_*}a_i(\tau)\left[c^\dagger\star D^{(0)}_ic-(D^{(0)}_ic)^\dagger\star c\right]\,,\,D^{(0)}_ic=\partial_ic-i\delta A_i\star c\,,
\end{align}
plus a diamagnetic term, which will not play a role here. We may rewrite this coupling in terms of the momentum operator, $p^{\mathrm{CF}}_i=-iD_i^{(0)}$, which is the same as the operator we introduced in Section \ref{sec: single particle}. Then we may write 
\begin{align}
\label{eq: vertex}
\Gamma_{ac^\dagger c}&=-\int d^2\mathbf{x}d\tau\, a_i(\tau)\,c^\dagger\,\frac{p^{\mathrm{CF}}_i}{m_*}\,c\,.
\end{align}

We work with a complete basis of non-commutative Landau level eigenstates, which we denote $\{|M,\alpha\rangle\}$, where $M$ is a Landau level index, and $\alpha$ parameterizes the Landau level degeneracy. The fermion propagator may then be written as
\begin{align}
G_{M}(\omega)&=\sum_\alpha\frac{|M,\alpha\rangle\langle M,\alpha|}{i\omega-E_M}\,,
\end{align}
where $E_M=\epsilon_M-\mu$, $\mu$ is a chemical potential that fixes the density to $\bar\rho$, and $\epsilon_M$ are the Landau level energies, Eq. \eqref{eq: LL energies}. Using this form for the propagator, along with the vertex in Eq. \eqref{eq: vertex}, the polarization tensor may be expressed as
\begin{align}
\pi_{xy}(\omega)&=\frac{1}{\mathrm{Area}}\int\frac{d\Omega}{2\pi}\sum_{M,N}\sum_{\alpha,\beta}\frac{\langle M,\alpha|\frac{p^{\mathrm{CF}}_x}{m_*}|N,\beta\rangle\langle N,\beta|\frac{p_y^{\mathrm{CF}}}{m_*}|M,\alpha\rangle}{[i(\omega+\Omega)-E_M][i\Omega-E_N]}\,\\
&=\frac{i\omega}{\mathrm{Area}}\sum_{M\neq N}\sum_{\alpha,\beta}\frac{\langle M,\alpha|\frac{p^{\mathrm{CF}}_x}{m_*}|N,\beta\rangle\langle N,\beta|\frac{p_y^{\mathrm{CF}}}{m_*}|M,\alpha\rangle}{(E_M-E_N)^2}(\Theta(-E_M)-\Theta(-E_N))+\mathcal{O}(\omega^2)\,.
\end{align}
We now use the fact 
\begin{align}
\frac{p^{\mathrm{CF}}_i}{m_\star}&=\frac{1}{i}[R^v_i,H]\,,
\end{align}
where here $R^v_i$ is defined (in mean field theory) to act as $R^v_ic=c\star x_i$. This allows us to rewrite,
\begin{align}
\pi_{xy}(\omega)&=\frac{i\omega}{\mathrm{Area}}\sum_{M\neq N}\sum_{\alpha,\beta}\langle M,\alpha|R_x^v|N,\beta\rangle\langle N,\beta|R^v_y|M,\alpha\rangle(\Theta(-E_M)-\Theta(-E_N))\,.
\end{align}

We now define the Landau level projection operators,
\begin{align}
\mathcal{P}_M=\sum_\alpha |M,\alpha\rangle\langle M,\alpha|\,,
\end{align}
such that
\begin{align}
\pi_{xy}(\omega)&=\frac{i\omega}{\mathrm{Area}}\sum_{M}\Tr_{\alpha}\left[\mathcal{P}_MR^v_x(1-\mathcal{P}_M)R^v_y-\mathcal{P}_MR^v_y(1-\mathcal{P}_M)R^v_x\right]\Theta(-E_M)\,,\\
&=\frac{i\omega}{\mathrm{Area}}\sum_M\Tr_\alpha\left[\mathcal{P}_M[R^v_x,R^v_y]+\mathcal{P}_MR^v_y\mathcal{P}_M R^v_x-\mathcal{P}_MR^v_x\mathcal{P}_M R^v_y\right]\,.
\end{align}
Here $\Tr_\alpha$ is the trace over the degenerate indices. To compute the projected vortex coordinates, $\mathcal{P}_MR_i^v\mathcal{P}_M$, we recall $p^{\mathrm{CF}}_i=\frac{1}{\ell_B^2}\varepsilon_{ij}(R^v_j-R_j)$, and we introduce operators, 
\begin{align}
r^{\mathrm{CF}}_i&=\frac{1}{2}(R^v_i+R_i)\,,\\ 
\tilde{p}_i^{\mathrm{CF}}&=\frac{2\ell_B^2}{\ell_{B+\delta \hat{B}}^2+\ell_B^2}p_i^{\mathrm{CF}}\,,\\
\mathcal{D}_i&=\frac{1}{1-\theta\delta B_{\mathrm{CF}}}(\tilde{p}_i^{\mathrm{CF}}-\delta B_{\mathrm{CF}}\varepsilon_{ij}r^{\mathrm{CF}}_j)\,,\\\theta&=\frac{1}{4}(\ell_B^2-\ell_{B+\delta \hat{B}}^2)\,,\,\delta B_{\mathrm{CF}}=4\frac{\ell_B^2-\ell^2_{B+\delta\hat{B}}}{(\ell^2_{B+\delta\hat{B}}+\ell_B^2)^2}\,,
\end{align}
which satisfy the same algebra as the corresponding operators in Section \ref{sec: single particle}. Importantly, $\mathcal{D}_i$ commutes with $p_i^{\mathrm{CF}}$ and therefore the Hamiltonian, meaning that it survives Landau level projection. As discussed in Section \ref{sec: single particle}, it also satisfies the algebra,
\begin{align}
[\mathcal{D}_x,\mathcal{D}_y]&=-i\delta \hat{B}\,.
\end{align}
In terms of these operators, we can express $R^v_i$ as
\begin{align}
R^v_i=\varepsilon_{ij}\left[\frac{1-\theta\delta B_{\mathrm{CF}}}{\delta B_{\mathrm{CF}}}\mathcal{D}_j-\frac{1}{\delta B_{\mathrm{CF}}}\left(1+\frac{\ell^2_{B}-\ell_{B+\delta \hat{B}}^2}{\ell_B^2+\ell_{B+\delta\hat{B}}^2}\right)\tilde{p}_j^{\mathrm{CF}}\right]\,.
\end{align}
Because the creation and annihilation operators are built out of $p^{\mathrm{CF}}$, the second term vanishes on projection, i.e.
\begin{align}
\mathcal{P}_M R^v_i\mathcal{P}_M=\frac{1-\theta\delta B_{\mathrm{CF}}}{\delta B_{\mathrm{CF}}}\varepsilon_{ij}\mathcal{D}_j=\frac{1}{\delta \hat{B}}\,\varepsilon_{ij}\mathcal{D}_j\,.
\end{align}
Thus, since $\mathcal{D}$ commutes with $p^{\mathrm{CF}}$ and the degeneracy of each Landau level is $\frac{\delta\hat{B}\times \mathrm{Area}}{2\pi}$, if $p$ Landau levels are filled,
\begin{align}
\pi_{xy}(\omega)&=\frac{i\omega}{\mathrm{Area}}\sum_M\Tr_\alpha\left[[R^v_x,R^v_y]-\frac{1}{\delta \hat{B}^2}[\mathcal{D}_x,\mathcal{D}_y]\right]\Theta(-E_M)\\
&=-\frac{\omega}{\mathrm{Area}}\times p\times\frac{\delta\hat{B}\times\mathrm{Area}}{2\pi}\times\left(\ell_B^2+\frac{1}{\delta\hat{B}}\right)\,.
\end{align}
But, due to the constraint, the composite fermion density is fixed to $\bar\rho=\frac{B}{2\pi}$, so $\delta\hat{B}\ell_B^2=1/p$, and this result becomes
\begin{align}
\pi_{xy}(\omega)&=-\frac{\omega}{2\pi}(p+1)\,.
\end{align}
The resulting Chern-Simons effective action is therefore
\begin{align}
S_{\mathrm{eff}}&=-\int d^2\mathbf{x}d\tau\,\frac{i(p+1)}{4\pi}ada+\mathcal{O}(\Theta)\,,
\end{align}
reflecting the correct topological order for the $\nu=p/(p+1)$ Jain state. Notably, the constraint has played an essential role in generating a properly quantized Chern-Simons level. 

We now comment on the background and mutual Chern-Simons terms. For a spatially uniform background probe field, $A_i'(\tau)$, the coupling to the composite fermions is the same as for $a_i(\tau)$, as in Eq. \eqref{eq: vertex}. The calculation of these terms is therefore identical, and their coefficients are also $p+1$. However, one should be careful in the interpretation of this result, since $\partial_0 A'_i$ is not the exactly the physical electric field, as discussed in Section \ref{sec: linear response}. Indeed, $A'$ does not even transform covariantly under left-acting $U(1)$ gauge transformations. Instead, it satisfies the modified infinitesimal transformation law,
\begin{align}
A'_i\rightarrow  A'_i+i\,\partial_i\lambda+i[\lambda,\delta A_i+A'_i]_\star\,.
\end{align}
This means that the Seiberg-Witten map cannot be applied directly to the probe, $A'_i$, and it must be modified accordingly. Rather than doing so here, particularly given the difficulties with constructing a gauge invariant mutual Chern-Simons term, we leave this to future work. Instead, we emphasize the physically transparent derivation of the Hall conductivity in Section \ref{sec: linear response}, which should be consistent with such an analysis.

\nocite{apsrev41Control}
\bibliographystyle{apsrev4-1}
\bibliography{LLL}

\end{document}